\documentclass{jfm}
\usepackage{graphicx,epsfig,subfig}
\usepackage{epstopdf}
\usepackage{natbib}
\usepackage{amsmath}
\usepackage{mathtools}
\usepackage{amssymb}
\usepackage{color}
\usepackage{setspace}
\usepackage{mathrsfs}
\usepackage[usenames,dvipsnames]{xcolor}
\usepackage[linesnumbered,ruled]{algorithm2e}

\newcommand\dd{\mathrm{d}}
\newcommand\zi{\mathrm{i}}
\newcommand{\jump}[1]{\ensuremath{[\![#1]\!]}}

\def\bdot{\raise.2em\hbox to .15em{.}}

\definecolor{gray}{gray}{0.8}
\definecolor{darkgray}{gray}{0.5}

\def\bdotblack{\raise.25em\hbox to .15em{.}}

\shorttitle{Wavy viscoelastic channel flow}
\title{Vorticity amplification in viscoelastic channel flows with long-wave surface distortions}

\shortauthor{J. Page \& T. A. Zaki}
\author{Jacob Page$^1$\thanks{jacob.page@ed.ac.uk} \& Tamer A. Zaki$^2$\thanks{t.zaki@jhu.edu}} 

\affiliation{$^1$School of Mathematics, University of Edinburgh, EH9 3FD, UK\\
$^2$Department of Mechanical Engineering, Johns Hopkins University, Baltimore, MD 21218, USA}

\pubyear{2021}
\volume{}
\pagerange{}
\date{\today}
\begin{document}

\maketitle

\begin{abstract}
    Surface distortions to an otherwise planar channel flow introduce vorticity perturbations. 
    We examine this scenario in viscoelastic fluids, and identify new mechanisms by which significant vorticity perturbations can be generated in both inertialess and elasto-inertial channel flows. 
    We focus on the case where the lengthscale of the surface distortion is much longer than the channel depth, where we find significant departure from plane shear (Page \& Zaki, \emph{J. Fluid Mech.} \textbf{801} 2016) due to the non-monotonic base-flow streamwise-normal elastic stress. 
    In inertialess flows, a purely elastic response results in streamlines deforming to match the bottom topography in the lower half the channel. 
    However, the vanishing stress at the centreline introduces a blocking effect,
    and the associated $O(1)$ jump in normal velocity is balanced by a narrow, large amplitude streamwise-oscillating `jet’, resulting in a localised, chevron-shaped vorticity perturbation field. 
    In elasto-inertial flows, resonance between the frequency of elasto-inertial `Alfven’ waves and the frequency apparent to an observer moving with the fluid results in vorticity amplification in a pair of critical layers on either side of the channel. 
    The vorticity in both layers is equal in magnitude and as such the perturbation vorticity field penetrates the full channel depth even when inertia is dominant.
    The results demonstrate that long-wave distortions -- which are relatively innocuous in Newtonian fluids -- can drive a significant flow distortion in viscoelastic fluids for a wide range of parameter values.
\end{abstract}


\section{Introduction}
\label{sec:intro}
Viscoelastic shear flows exhibit a range of chaotic dynamics depending on the flow parameters, from inertialess elastic turbulence \citep{Groisman2000} to a modified, drag-reduced, quasi-Newtonian turbulence when inertia dominates \citep{White2008}.
Between these two extremes, a further chaotic flow state -- elasto-inertial turbulence (EIT) -- can be sustained \citep{Samanta2013,Dubief2013}. 
The area of the parameter space where EIT is found overlaps with various exact coherent states which connect to either Newtonian Tollmien-Schlichting (TS) waves \citep{Lee2017,Shekar2020} or to a newly discovered centre-mode instability \citep{Page2020,Garg2018}. 
This raises the intriguing question as to whether the self-sustaining mechanism of EIT is rooted in purely Newtonian or elastic dynamics, given recent evidence that the centre mode instability persists in the inertialess limit for extremely dilute solutions \citep{Khalid2021,Buza2021}. 

In planar flows, both elastic turbulence and EIT exist at subcritical parameter settings, which is in agreement with the known characteristics of the finite-amplitude travelling wave solutions which connect to TS waves and the centre mode \citep{Shekar2020,Page2020,Buza2021}. 
Earlier work has also sought to connect elastic turbulence to linear instabilities in other configurations with streamline curvature in the basic state \citep[e.g. Taylor-Couette flow][]{Shaqfeh1996}.
In this scenario, elastic turbulence in parallel flows would always require a finite amplitude perturbation since these instabilities vanish in the absence of an elastic `hoop' stress \citep{Meulenbroek2004,Morozov2005,Pan2013}.
Inertialess viscoelastic flows also feature new linear transient growth mechanisms, the most significant being an inertialess version of `lift-up' driven by polymer forces \citep{Jovanovic2010,Jovanovic2011}.
Given this subcritical picture, we focus here on a problem related to receptivity: how is streamline curvature in the bulk of a channel flow induced by surface roughness at the boundaries?
We identify new mechanisms by which significant vorticity fluctuations can be established in the flow in both purely elastic and elasto-inertial regimes. 
Beyond the inherent interest in these flow patterns as a platform for secondary instability, we believe that the amplification mechanisms and associated asymptotic solutions will be of use in the study of the aforementioned instability waves, where the underlying mechanisms are unknown.

In earlier work \citep{Page2016} we studied the analogous problem in a Couette geometry, a viscoelastic version of the Newtonian analysis by \citet{charru2000}. 
In viscoelastic Couette flow, the vortical response to a monochromatic boundary perturbation depends on the value of two dimensionless parameters: $\alpha$, the dimensionless wavenumber of the surface distortions, and $\Sigma$, the ratio of a viscoelastic critical layer depth to the channel height.
The viscoelastic critical-layer depth is the point at which the phase speed of (backward) propagating elasto-inertial waves matches the base-flow velocity. 
These elasto-inertial waves have a speed proportional to the square root of the streamwise normal elastic stress, analogous to Alfven waves in magnetohydrodynamics \citep{chandrasekhar}. 
Across the layers in the flow where the base-flow speed matches the elastic wave speed the equations change type \citep[a transition from sub- to supersonic, see][]{Yoo1985}; at these points a resonance exists between the apparent frequency of oscillation of the wall to an observer moving with the base flow and the frequency associated with the elastic waves.  
This resonance results in a significant vorticity amplification in a thin region \citep{Page2016} driven by a kinematic reverse-Orr mechanism in the polymer torque \citep{Page2015}, and the location of this layer delineates three regimes of vortical response in the Couette flow:
(i) In shallow elastic flow the channel depth is small compared to the roughness lengthscale, and the critical layer is outside of the flow domain; vortical perturbations fill the channel and the vorticity amplifies at the top wall.
(ii) In elasto-inertial flows \citep[referred to as `transcritical' in ][]{Page2016}, the elastic critical layer is inside the flow domain and within a wavelength of the lower wall, leading to large vorticity amplification at that location via the resonance mechanism.
(iii) In deep elastic flows the critical layer is far from the lower wall and the vorticity perturbation decays monotonically with height. 

The three flow regimes were subsequently confirmed in experiments with real polymer solutions \citep{Haward2017,Haward2018a,Haward2018b} in a pressure-driven channel flow. 
In deeper channels (short wavelength wall disturbances), there is a near-exact correspondence with the Couette case, since on the lengthscale of the roughness the base flow looks like simple shear.
However, in shallow channels --notably where experimental data is challenging to obtain-- this will no longer be the case since the perturbation will feel the impact of the non-monotonic background velocity profile at leading order, even in the absence of inertia (where the Newtonian response would be insensitive to background velocity profile). 
These deviations are driven largely by the base streamwise normal polymer stress which is proportional to the local shear rate and hence vanishes at the centreline. This base flow stress provides a mechanism for elasto-inertial wave propagation in flows with inertia and leads to a purely elastic response otherwise.  
Therefore, in shallow channels we must account for the fact that (i) there will be \emph{two} critical layers in the flow domain where the base velocity matches the elastic wave speed and (ii) the vanishing elastic stresses at the centreline may disrupt the dominant balance that leads to vorticity amplification in the shallow elastic Couette flow. 
In this paper we will examine both of these effects in detail, where we will see that both provide new mechanisms for significant vorticity amplification in shallow channels.

The remainder of this paper is structure as follows:
in \S\ref{sec:setup_local} we introduce the governing equations and show numerically the new regimes in the flow over a long Gaussian bump.
In \S\ref{sec:asymptotics} we derive matched asymptotic expansions for the two new behaviours which appear in a viscoelastic channel under the joint assumption of long-wave disturbances and high Weissenberg numbers; we also discuss the amplification mechanisms.
Finally, conclusions are provided in \S\ref{sec:conclusions}.

\section{Vortical response to wall roughness in viscoelastic channel flows}
\label{sec:setup_local}
\subsection{Setup}
\begin{figure}
    \centering
    \hspace{15mm}\includegraphics[width=0.5\textwidth]{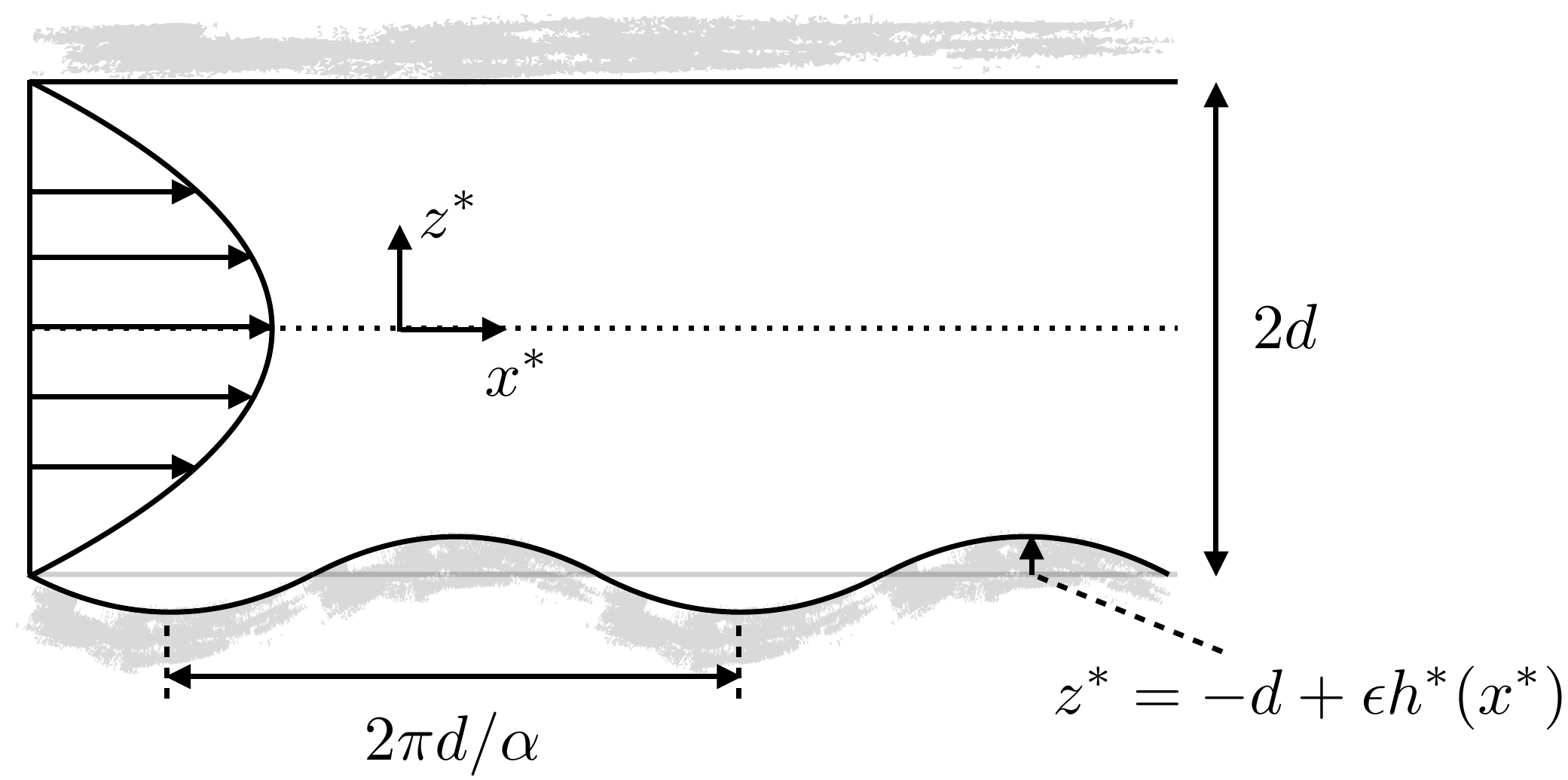}
    \caption{Schematic of the flow configuration considered in this paper, with variables shown in their dimensional form. The problem is non-dimensionalized by the channel half-height, $d$ and the wall shear rate, $\dot{\gamma}$, of the base-flow velocity. A monochromatic surface topography is shown for illustration.}
    \label{fig:schematic}
\end{figure}
We consider the pressure-driven, steady flow of an Oldroyd-B fluid in a two-dimensional, planar channel. 
The governing equations are 
\begin{subequations}
\begin{align}
    \boldsymbol \nabla \cdot \boldsymbol u &= 0, \\
    \boldsymbol u \cdot \boldsymbol \nabla \boldsymbol u &= -\boldsymbol \nabla p + \frac{\beta}{R}\nabla^2\boldsymbol u
    + \frac{(1-\beta)}{R}\boldsymbol \nabla \cdot \mathbf T, \\
    \boldsymbol u \cdot \boldsymbol \nabla \mathbf C + \mathbf T &= \mathbf C \cdot \boldsymbol \nabla \boldsymbol u + (\boldsymbol \nabla \boldsymbol u)^\top \cdot \mathbf C,
\end{align}
\label{eqn:full_governing}
\end{subequations}
where $\mathbf T = W^{-1}(\mathbf C - \mathbf I)$ relates the polymeric stress and conformation tensor. 
The equations have been non-dimensionalised by the wall shear rate, $\dot{\gamma}$, and channel half-height, $d$, which defines the Reynolds and Weissenberg numbers as $R:=\dot{\gamma} d^2/\nu$ and $W:=\dot{\gamma}\varsigma$ respectively, where $\varsigma$ is the relaxation time of the polymer chains and $\nu$ is the total kinematic viscosity of the fluid; the parameter $\beta := \nu_s/\nu$ is the ratio of solvent-to-total viscosity. 

In the absence of fluctuations, the streamwise-independent solution of (\ref{eqn:full_governing}) is the standard Poiseuille velocity profile with associated polymeric stresses
\begin{equation}
    U(z) = \frac{1}{2}(1-z^2), \quad T_{11}(z) = 2W(U')^2 = 2Wz^2, \quad T_{13}(z) = U' = -z,
\end{equation}
where $z\in [-1,1]$ is the vertical coordinate, and primes denote derivatives of base state variables. 
The linear scaling of the streamwise normal stress, $T_{11}$, with $W$ is the leading cause of deviations from a standard ``Newtonian'' response, both in the channel configuration considered here and in the simpler Couette configuration \citep{Page2016}. 

We are interested in the steady perturbation field induced by the addition of a small amplitude surface roughness to the lower wall of the channel, $z=-1+\epsilon h(x)$, where $\epsilon \ll 1$.
A schematic of the configuration is shown in figure \ref{fig:schematic}. 
Non-trivial vortical perturbations arise at $O(\epsilon)$ via a slip condition on the perturbation velocity at the lower wall, $u(x,z=-1) = -h(x)$ (in dimensional variables, $u^*(x^*,z^* = -d) = -\dot{\gamma} h^*(x)$), which results from the requirement that the total velocity vanishes on the solid boundary. The no-penetration condition, $w(x,z=-1)=0$, is unchanged in the linearised problem.
Assuming that the surface roughness is a smooth function of $x$, we write it as a Fourier series 
\begin{equation*}
    h(x) = \sum_\alpha \hat{h}_{\alpha} e^{\zi \alpha x},
\end{equation*}
and solve for the monochromatic linear response to each individual surface wavenumber $\alpha$,
\begin{subequations}
    \begin{align}
        \zi \alpha \hat{u} + D\hat{w} = 0, \\
        \zi \alpha U\hat{u} + \hat{w}U' = -\zi \alpha \hat{p} + 
        \frac{\beta}{R}\left(D^2 - \alpha^2\right)\hat{u} + \frac{(1-\beta)}{R}\left(\zi \alpha \hat{\tau}_{11} +D\hat{\tau}_{13}\right), \\
        \zi \alpha U\hat{w} = -D\hat{p} +        
        \frac{\beta}{R}\left(D^2 - \alpha^2\right)\hat{w} + \frac{(1-\beta)}{R}\left(\zi \alpha \hat{\tau}_{13} +D\hat{\tau}_{33}\right), \\
        \zi \alpha U\hat{\tau}_{11}+\hat{w}T_{11}' + \frac{1}{W}\hat{\tau}_{11} = 
        2\zi \alpha T_{11}\hat{u} + 2T_{13}D\hat{u} + 2U'\hat{\tau}_{13} +\frac{2\zi \alpha}{W}\hat{u}, \\
        \zi \alpha U\hat{\tau}_{13}+\hat{w}T_{13}' + \frac{1}{W}\hat{\tau}_{13} = 
        \zi \alpha T_{11}\hat{w} + U'\hat{\tau}_{33} +\frac{1}{W}\left(\zi \alpha \hat{w} + D\hat{u}\right), \\
        \zi \alpha U\hat{\tau}_{33} + \frac{1}{W}\hat{\tau}_{33} = 
        2\zi \alpha T_{13}\hat{w} +\frac{2}{W}D\hat{u}.
    \end{align}
    \label{eqn:linear_full}
\end{subequations}
where the slip condition on the full velocity is scaled appropriately for each component wave, $\hat{u}(z=-1) = -\hat{h}_{\alpha}$, and $D$ is the wall-normal derivative. 

We are interested in both the response to localised roughness, which we will explore numerically, and the vortical perturbations induced by monochromatic surface waves, which we will examine with matched asymptotic expansions.
For the numerics, we solve equations (\ref{eqn:linear_full}) --and other similar systems appearing in this paper-- using an expansion in Chebyshev polynomials in $z$. Typically $N_c\approx 200$ polynomials is sufficient. 

\subsection{Response to local roughness}
\begin{figure}
    \centering
    \includegraphics[width=\textwidth]{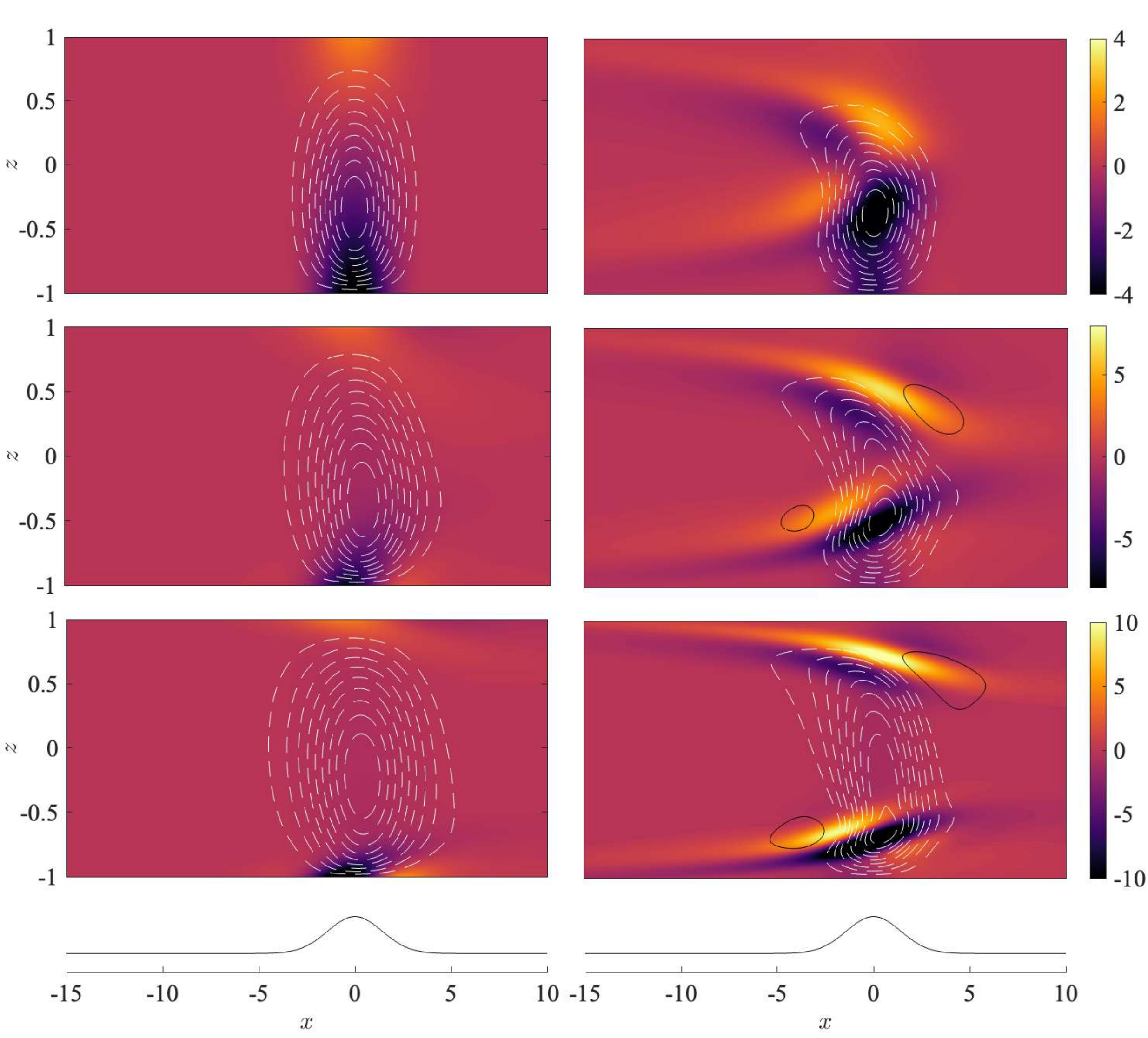}
    \caption{Response to Gaussian bump with $l_x=2$ in Newtonian (left) and viscoelastic (right) flows with $W=200$ and $\beta=0.5$. The Reynolds number is matched between the Newtonian and viscoelastic calculations and increases from top to bottom, $R=\{1, 200, 1000\}$. Colours show the spanwise vorticity perturbation, lines are the perturbation streamfunction (solid black positive, dashed white negative).}
    \label{fig:local_bump}
\end{figure}
In the Couette configuration studied by \citet{Page2016}, significant vorticity amplification at high-$W$ occurred either at the top wall or at a critical layer in the bulk of the flow (where the base flow speed is equal to the elastic wave speed; see discussion in \S\ref{sec:intro}). 
New mechanisms for vorticity amplification are possible in the channel flow studied here, and will be shown to be associated with the non-monotonic base-flow velocity profile and the variation of the normal stress $T_{11}$ with depth $z$. 
Intuitively, these differences only manifest in shallow channels, where the characteristic lengthscale of the roughness, $l_x^*$, is comparable to or longer than the channel half-height, $d$. 
Small-scale roughness sees a background flow that locally is a close approximation to simple shear, and the corresponding Couette behaviour is recovered for the vorticity field. 

The new regimes of vorticity amplification in a viscoelastic channel flow are summarised in figure \ref{fig:local_bump}, where the spanwise vorticity perturbations induced by a long Gaussian bump, $h(x)=\text{exp}(-x^2 / l_x^2)$, where $l_x:=l_x^*/d=2$, are examined for a particular set of viscoelastic parameters and three Reynolds numbers, $R\in\{1, 200, 2000\}$, alongside the response in a Newtonian fluid. 

For all three values of $R$ in the Newtonian fluid, the response is a straightforward modification of the response in Couette flow to a channel geometry:
At low-$R$, the perturbation field is a Stokes flow solution and is independent of the details of the background flow due to the absence of advection -- a response which is identical to ``shallow viscous'' flow in a simple shear \citep{charru2000}.
At higher Reynolds numbers, a shallow-channel version of the ``inviscid'' regime of \citet{charru2000} is found, where the vorticity response at the lower wall is confined to a thin layer of thickness $\delta \sim R^{-1/3}$. 
There is a larger, irrotational flow response which fills the domain, and a weak vorticity perturbation is established in a thin wall layer around $z=1$ by adjustment to the no-slip boundary condition.

In contrast, the viscoelastic flow response is strikingly different to both the Newtonian flow fields and the response in a viscoelastic Couette flow at the same parameter settings for all three values of $R$. 
In the near-inertialess flow at $R=1$, the vorticity is amplified around the channel centreline directly above the bump, with the elongated stripes of vorticity extending significantly upstream of the bump. 
This \emph{shallow elastic} behaviour is markedly different from the same regime in Couette flow, where vorticity amplification occurs at the top wall.

At the largest value of $R$, the perturbation vorticity is amplified in stripes which sit some distance from the lower wall. 
This \emph{shallow elasto-inertial} response is familiar from the Couette configuration, where vorticity is amplified in a viscoelastic critical layer at which the base-flow velocity matches the elastic wave speed. 
In contrast to that behaviour, vorticity perturbations of a similar magnitude are also established on the opposite, smooth-walled side of the channel, presumably at a second critical layer. 
Therefore, while the penetration depth of the vorticity in an elasto-inertial Couette flow is proportional to the critical layer depth, here the vortical perturbations fill the channel. 
Finally, the intermediate value of $R=200$ shows a mixture of these two behaviours and occurs in a transition between these two distinct regimes. 

In all cases, the vorticity is amplified relative to its Newtonian counterpart, and our goal now is to identify the mechaninics underpinning each of the new regimes by the construction of asymptotic solutions of the flow response to a single wavenumber $\alpha$ in the long-wave ($\alpha\ll 1$) but high Weissenberg ($\alpha W\gg 1$) limit. 
In both cases we will identify the relevant physical mechanisms and use the solutions to estimate the level of vorticity amplification in each flow type.
We will also discuss the transition between the two regimes, which can be understood in terms of the dependence of the $z$-location(s) of the elasto-inertial critical layers on the fluid elasticity, $W/R$.

\section{Asymptotics for monochromatic wall roughness}
\label{sec:asymptotics}
\begin{figure}
    \centering
    \includegraphics[width=\textwidth]{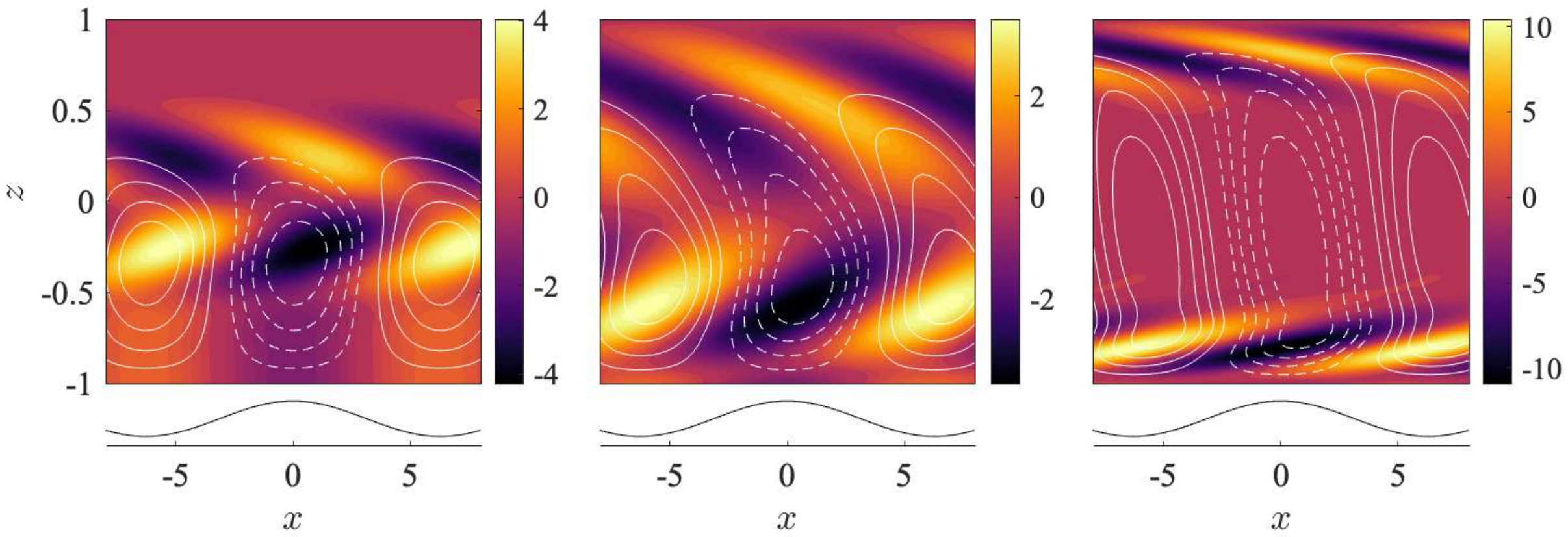}
    \caption{Vortical response to monochromatic surface roughness $h(x) = \text{cos}\,\alpha x$, with $\alpha=0.5$. (Left) $W=200$, $\beta=0.2$, $R=1$; (centre) $\beta=0.5$, $R=100$, $W=100$; (right) $W=200$, $\beta=0.5$, $R=2000$. Colours are spanwise vorticity perturbations, lines the perturbation streamfunction.}
    \label{fig:mono_regimes}
\end{figure}
Both the shallow elastic and shallow elasto-inertial behaviours --and an intermediate state-- are recovered in figure \ref{fig:mono_regimes} for monochromatic wall roughness, $h(x) = \text{cos}\,\alpha x$. 
In the shallow elastic case, the flow response is clearly confined to the lower part of the channel, with vorticity amplifying in a chevron pattern about $z=0$. 
In the shallow elasto-inertial regime, vorticity is generated in two rows of tilted stripes, on either side of the channel and aligned with the background shear. 

In this section we will construct asymptotic solutions for both the shallow elastic and shallow elasto-inertial regimes. 
We will first derive the leading order equations in the long-wave limit, which are slightly different between the two cases owing to the exclusion or inclusion of inertia, before finding approximate solutions of these systems in the (singular) high Weissenberg number limit via matched asymptotic expansions. 

\subsection{The shallow elastic regime}
\label{sec:shallow_elastic}
The shallow elastic regime is associated with significant vorticity amplification at the channel centreline. 
We have seen this response occurs at high elasticity, i.e. when $E=W/R \gg 1$ (a more careful discussion of when this regime can be expected is provided in \S\ref{sec:discussion}). 
Assuming $\alpha \ll 1$, we adopt a long-wave scaling,
\begin{subequations}
    \begin{align}
        \hat{u} = u, \quad
        \hat{w} &= \alpha w, \quad
        \hat{p} = Wp, \\
        \hat{\tau}_{11} = W\tau_{11}, \quad 
        \hat{\tau}_{13} &= \alpha W\tau_{13}, \quad
        \hat{\tau}_{33} = \alpha \tau_{33},
    \end{align}
\end{subequations}
where we have assumed that the pressure scales with the viscoelastic stresses on account of the large elasticity in this regime.
At leading order in $\alpha$ our equations are,
\begin{subequations}
    \begin{align}
        \zi u + Dw = 0, \\
        0 = -\zi p + 
        \frac{\varepsilon \beta}{R}D^2u + \frac{(1-\beta)}{R}\left(\zi \tau_{11} +D\tau_{13}\right), \\
        0 = -Dp, \\
        \zi U\tau_{11}+wA_{11}' + \varepsilon\tau_{11} = 
        2\zi A_{11}u + 2\varepsilon T_{13}Du + 2U'\tau_{13}, \\
        \zi U\tau_{13}+\varepsilon w T_{13}' + \varepsilon\tau_{13} = 
        \zi A_{11}w + \varepsilon U'\tau_{33} +\varepsilon^2 Du, \\
        \zi U\tau_{33} + \varepsilon \tau_{33} = 
        2\zi T_{13}w + 2\varepsilon Dw,
    \end{align}
    \label{eqn:se_full}
\end{subequations}
where we have introduced $A_{11} := T_{11}/W = O(1)$ as the scaled streamwise normal base stress and $\varepsilon:=1/(\alpha W)$, which we will subsequently take to be our small parameter. 

\begin{figure}
    \centering
    \includegraphics[width=\textwidth]{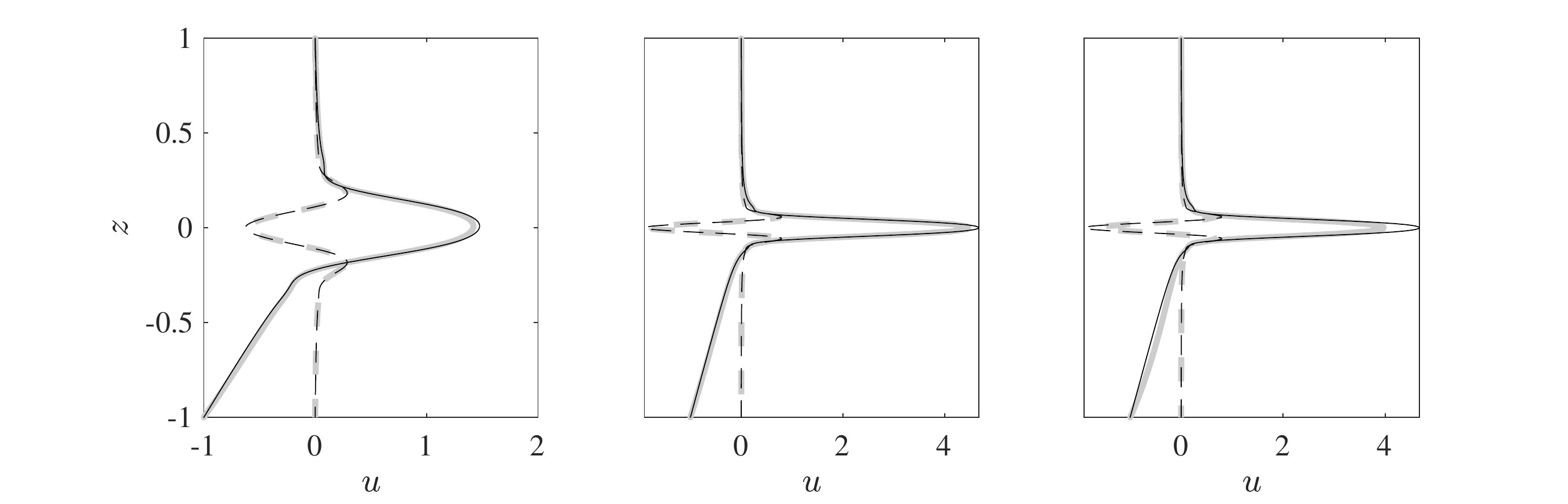}
    \includegraphics[width=\textwidth]{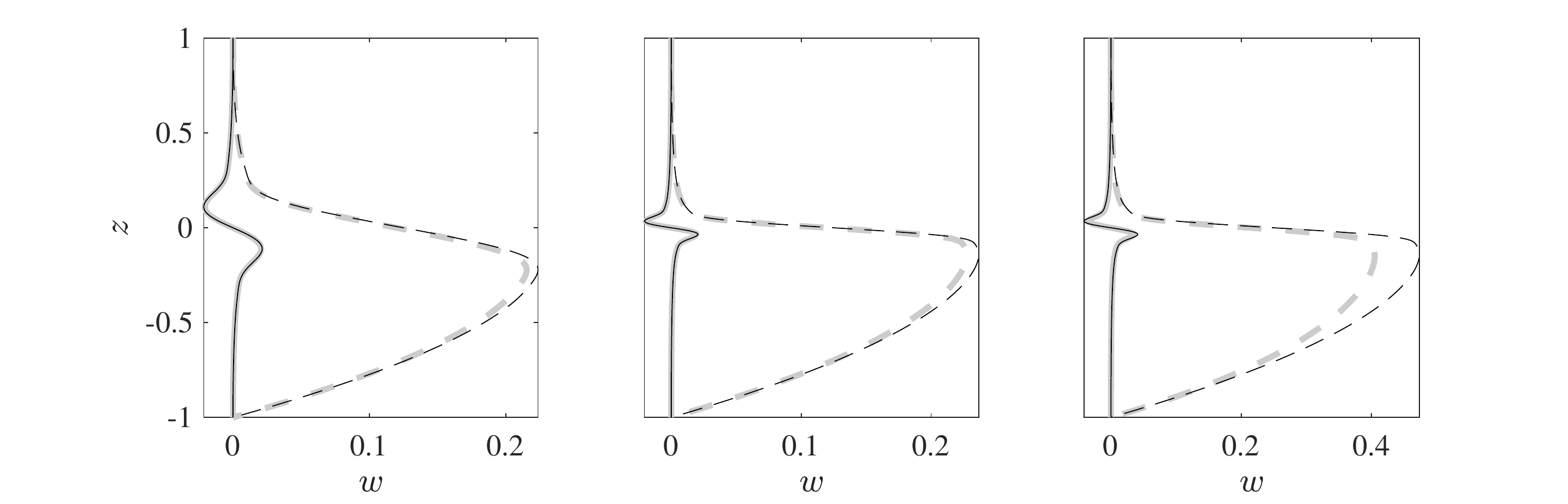}
    \caption{Comparison of full equations (grey) and shallow elastic approximation, with $\alpha=0.5$, $\varepsilon=10^{-3}$ (left), $\alpha=0.5$, $\varepsilon=10^{-5}$ (centre) and $\alpha=1$, $\varepsilon=10^{-5}$ (right). For all cases we have set $\beta=0.2$ and $R=1$. Note the good agreement even at moderate $\alpha=1$. Solid and dashed lines indicate the real and imaginary components of the solution respectively.}
    \label{fig:se_examples}
\end{figure}
The numerical solution of the shallow elastic system (\ref{eqn:se_full}) is compared to the full numerical solution of the governing equations for two moderate values of $\alpha$ in figure \ref{fig:se_examples}.
The agreement between the true velocities and those found from the long-wave system remain relatively accurate even when $\alpha=1$. 
Notably, the amplitude in the `jet' of $u$ seen at the centreline, which is related to the spanwise vorticity amplification observed in figures \ref{fig:local_bump} and \ref{fig:mono_regimes}, is fixed by the value of $\varepsilon$.
For very small $\varepsilon$, the vertical velocity response is increasingly confined to the lower half of the channel, and the jump in $w$ across $z=0$ is balanced by the strong streamwise velocity fluctuations at this location. 
Given the excellent agreement between solutions of (\ref{eqn:se_full}) and the full system of equations (\ref{eqn:linear_full}), we will explore the mechanics of the shallow elastic regime by finding asymptotic solutions of (\ref{eqn:se_full}) in the limit $\varepsilon\ll 1$, and do not seek to compute higher order corrections in $\alpha$.

\emph{Outer solution:} At leading order in $\varepsilon$, the viscoelastic stresses from equation (\ref{eqn:se_full}) are
\begin{subequations}
    \begin{align}
        \tau_{11}^0 &= \zi A_{11}'\phi_0 + 2\zi A_{11}D\phi_0, \\
        \tau_{13}^0 &= A_{11}\phi_0,
    \end{align}
    \label{eqn:se_bulk_stress}
\end{subequations}
where we have introduced the variable $\phi:=w/U$, which is proportional to the streamline displacement (equal under a phase shift of $\pi/2$).
The terms contributing to the normal stress are due to (i) base state stress maintained on perturbed streamlines and (ii) compression (or expansion) of streamlines carrying base-state stress \citep{Rallison1995}. 
The polymer shear stress is due to tilting of base state streamlines.
Note that $\tau_{33}$ does not couple to the other equations at this order.

The streamwise momentum equation reduces to the simple requirement that the streamwise polymer force vanishes everywhere in the channel; 
no pressure perturbation is required since the perturbation velocity satisfies continuity automatically (see below).
The net-zero polymer force can be converted into a condition on the streamline displacement $\phi$,
\begin{align}
    0 &= \frac{(1-\beta)}{R}\left(\zi \tau_{11}^0 + D\tau_{13}^0\right), \nonumber \\
    &= -\frac{(1-\beta)}{R}A_{11}D\phi_0,
\end{align}
which simply requires that $\phi_0=\text{constant}$, except at $z=0$ where the base stress $A_{11}(z=0)=0$. 
The boundary conditions are that the streamline displacement match the lower wall topography at $z=-1$, $\phi_0(z=-1)=\zi h$, and vanish at the upper wall, $\phi_0(z=1)=0$, which results in the following solution,
\begin{subequations}
    \begin{align}
        \phi_0^+ &= 0, \\
        \phi_0^- &= \zi h,
    \end{align}
\end{subequations}
where $\phi^+ := \phi(z>0)$ and $\phi^- := \phi(z<0)$ respectively.
Therefore, the streamline displacement in the lower half of the channel is a purely elastic response which mimics the lower wall across the full half-depth of the channel, before the vanishing base state stress at $z=0$ provides a blocking effect resulting in unperturbed streamlines above.

The vertical velocity perturbations at leading order are generated by the tilting of the mean streamlines to match the lower wall topography, hence $w_0^-=\zi h U(z)$ in the lower half of the channel and $w_0^+=0$ above. 
The vertical velocity therefore experiences an $O(1)$ jump, $\jump{w_0} = -\zi h/2$, while the streamwise velocity is continuous, with $u_0^-=-hU'(z)$ and $u_0^+=0$. 
There is therefore a boundary layer of thickness $\delta$ at $z=0$, in which the $O(1)$ jump in vertical velocity must be balanced by a streamwise velocity perturbation of magnitude $\sim 1/\delta$. 
This streamwise velocity has associated with it a pressure perturbation of order $\delta$ which is constant across the channel depth (since $D p_j=0$ at all orders). 
Therefore, higher order corrections to the outer solution are required and the associated asymptotic expansion takes the form,
\begin{subequations}
        \begin{align}
            \hat{u} = u_0\; +\; &\delta u_1 + \cdots ,\\
            \hat{w} = w_0\; +\; &\delta w_1 + \cdots, \\
            \hat{p} = \;\hphantom{p_0+}  &\delta p_1 + \cdots, \\
            \hat{\tau}_{11} = \tau_{11}^0\; +\; &\delta \tau_{11}^1 + \cdots, \\
            \hat{\tau}_{13} = \tau_{13}^0\; +\; &\delta \tau_{13}^1 + \cdots, \\
            \hat{\tau}_{33} = \tau_{33}^0\; +\; &\delta \tau_{33}^1 + \cdots,
        \end{align}
\end{subequations}
where the boundary layer thickness has not yet been determined, but we assume $\delta \gg \varepsilon$.

At $O(\delta)$ we have,
\begin{subequations}
    \begin{align}
        \zi u_1 + D w_1 = 0, \\
        0 = -\zi p_1+ 
        \frac{(1-\beta)}{R}\left(\zi \tau_{11}^1 +D\tau_{13}^1\right), \\
        0 = -Dp_1, \\
        \zi U\tau^1_{11}+w_1 A_{11}' = 
        2\zi A_{11}u_1 + 2T_{13}Du_1 + 2U'\tau^1_{13}, \\
        \zi U\tau_{13}^1 = 
        \zi A_{11}w_1, \\
        \zi U\tau^1_{33} = 2\zi T_{13}w_1.
    \end{align}
    \label{eqn:bulk_first_order}
\end{subequations}
Again, the vertical normal stress does not couple back to the other variables.
Rearranging the equations for the polymer stresses to obtain $\tau_{11}^1$ and $\tau_{13}^1$ yields expressions which are identical in form to the leading order solution (\ref{eqn:se_bulk_stress}). The only difference from the leading order solution for streamline displacement is that the polymer force is now balanced by the pressure correction $p_1$,
\begin{equation}
    0 = -\zi p_1  - \frac{(1-\beta)}{R}A_{11}D\phi_1.
\label{eqn:fo_xmtum}
\end{equation}
The pressure $p_1$ is constant across the channel. 
The constant pressure gradient must be balanced by the polymer force which implies that $D\phi_1 \sim 1/z^2$ as $z\to 0$ -- 
the streamline displacement must increase as the centreline is approached to balance the decreasing base-state stress.
Solving for the correction to the streamline displacement, our expansion to $O(\delta)$ is,
\begin{subequations}
    \begin{align}
        \phi^+ &= \hphantom{ \zi h +}\;\; \delta\frac{\zi R p_1}{2(1-\beta)}\left(\frac{1}{z} - 1\right)+\cdots, \\
        \phi^- &= \zi h + \; \delta\frac{\zi R p_1}{2(1-\beta)}\left(\frac{1}{z} + 1\right)+\cdots.
    \end{align}
    \label{eqn:se_outer_phi}
\end{subequations}

\emph{Inner expansion:} 
We introduce an inner variable, $\eta:=z/\delta(\varepsilon)$, and considering the inner limit of the outer solution (\ref{eqn:se_outer_phi}) indicates that $u\sim 1/\delta$ when $\eta=O(1)$.
On the other hand, the polymer stresses (not shown) are weak near the centreline, with $\tau_{11}\sim \delta$ and $\tau_{13}\sim \delta^2$.
The full set of inner variables are defined as follows,
\begin{subequations}
    \begin{align}
        \overline{u} = \delta u, \quad
        \overline{w} &= w, \quad
        \overline{p} = p/\delta, \\
        \overline{\tau}_{11} = \tau_{11}/\delta, \quad
        \overline{\tau}_{13} &= \tau_{13}/\delta^2, \quad
        \overline{\tau}_{33} = \tau_{33}/\delta.
    \end{align}
    \label{eqn:inner_var_scales_se}
\end{subequations}
Applying this scaling in the streamwise momentum equation, a dominant balance between the polymer force, pressure gradient and solvent diffusion implies that $\delta = \varepsilon^{1/4}$. 

With the new scalings, the continuity and momentum equations read
\begin{subequations}
    \begin{align}
        \zi \overline{u} + \frac{\dd \overline{w}}{\dd \eta} = 0, \\
        0 = -\zi \overline{p} + 
        \frac{\beta}{R}\frac{\dd^2\overline{u}}{\dd \eta^2} + \frac{(1-\beta)}{R}\left(\zi \overline{\tau}_{11} +\frac{\dd \overline{\tau}_{13}}{\dd \eta}\right), \\
        0 = -\frac{\dd \overline{p}}{\dd \eta}, \\
        \frac{\zi}{2} \left(1-\varepsilon^{1/2}\eta^2\right)\overline{\tau}_{11}+ 4\eta\overline{w} + \varepsilon\overline{\tau}_{11} = 
        4\zi \eta^2 \overline{u} + 2\varepsilon^{1/2}\eta \frac{\dd \overline{u}}{\dd \eta} - 2\varepsilon^{1/2}\eta\overline{\tau}_{13}, \\
        \frac{\zi}{2} \left(1-\varepsilon^{1/2}\eta^2\right)\overline{\tau}_{13}-\varepsilon^{1/2}\overline{w} + \varepsilon\overline{\tau}_{13} = 
        2\zi \eta^2 \overline{w} - \varepsilon \eta\overline{\tau}_{33} +\varepsilon \frac{\dd \overline{u}}{\dd \eta}, \\
        \frac{\zi}{2} \left(1-\varepsilon^{1/2}\eta^2
        \right)\overline{\tau}_{33} + \varepsilon \overline{\tau}_{33} = 
        -2\zi \eta \overline{w} + 2\varepsilon^{1/2} \frac{\dd \overline{w}}{\dd \eta}.
    \end{align}
    \label{eqn:inner_all}
\end{subequations}
To leading order, the streamwise-normal and polymer shear stresses satisfy,
\begin{subequations}
    \begin{align}
        \frac{\zi}{2}\overline{\tau}_{11}^0 = 
        4\zi \eta^2 \overline{u}_0, \\
        \frac{\zi}{2}\overline{\tau}_{13}^0 = 
        2\zi \eta^2 \overline{w}_0.
    \end{align}
    \label{eqn:inner_stress}
\end{subequations}
The streamwise momentum equation can then be written as a second order equation for $\overline{u}_0$ with constant forcing,
\begin{equation}
    \frac{\dd^2 \overline{u}_0}{\dd \eta^2} + g(\beta) \eta^2 \overline{u}_0 = \frac{\zi R p_1}{\beta},
    \label{eqn:inner_u0}
\end{equation}
where $g(\beta):=4\zi(1-\beta)/\beta$ and we have used the fact that $\overline{p}_0 = p_1$.

An exact solution of equation (\ref{eqn:inner_u0}) is provided in Appendix \ref{sec:app_se} and can be expressed in terms of modified Bessel functions.
The far-field behaviour of this inner solution is
\begin{equation}
    \overline{u}_0(\eta \to \pm\infty) \sim \frac{R p_1}{4(1-\beta)}\left(\frac{1}{\eta^2} - \frac{6}{g\eta^6} + \cdots \right).
\end{equation}
which automatically matches with the outer solution for $u$ (not shown).
In fact, the full solution for $\overline{u}_0$ is directly proportional to the prsesure correction $p_1$ and can be written in the form
$\overline{u}_0(\eta) = p_1 \overline{\Theta}_0(\eta)$. 

To determine the dependence of $p_1$ on the wall amplitude $h$ we need to connect the boundary layer jet in the streamwise velocity to the jump in outer vertical velocity, which can be done by integrating the continuity equation and matching with the constant $w \sim \zi h/2$ in the bulk,
\begin{equation}
    \overline{w}_0(\eta) = \frac{\zi h}{2} - \zi \int_{-\infty}^{\eta}\overline{u}_0(\eta')\dd \eta'.
\end{equation}
The limiting form as $\eta \to \pm \infty$ can be determined by using the asymptotic approximation to $\overline{u}_0$ and integrating term-by-term, yielding
\begin{subequations}
\begin{align}
    \overline{w}_0(\eta \to -\infty) &\sim \frac{\zi h}{2} + \frac{\zi R p_1}{4(1-\beta)}\frac{1}{\eta} + \cdots, \quad \text{and} \\
    \overline{w}_0(\eta \to +\infty) &\sim \frac{\zi h}{2} - \zi p_1 \int_{-\infty}^{\infty}\overline{\Theta}_0(\eta')\dd \eta'  + \frac{\zi R p_1}{4(1-\beta)}\frac{1}{\eta} + \cdots.
\end{align}
\end{subequations}
Our inner solution for the vertical velocity therefore predicts a jump $\jump{\overline{w}_0} =  - \zi p_1 \int_{-\infty}^{\infty}\overline{\Theta}_0(\eta')\dd \eta'$, which must match the jump in the outer solution, $\jump{w_0}=-\zi h/2$, and implies
\begin{equation}
    p_1 = \frac{h}{2\int_{-\infty}^{\infty}\overline{\Theta}_0(\eta')\dd \eta'}.
\end{equation}
This completes the solution.

\begin{figure}
    \centering
    \includegraphics[width=\textwidth]{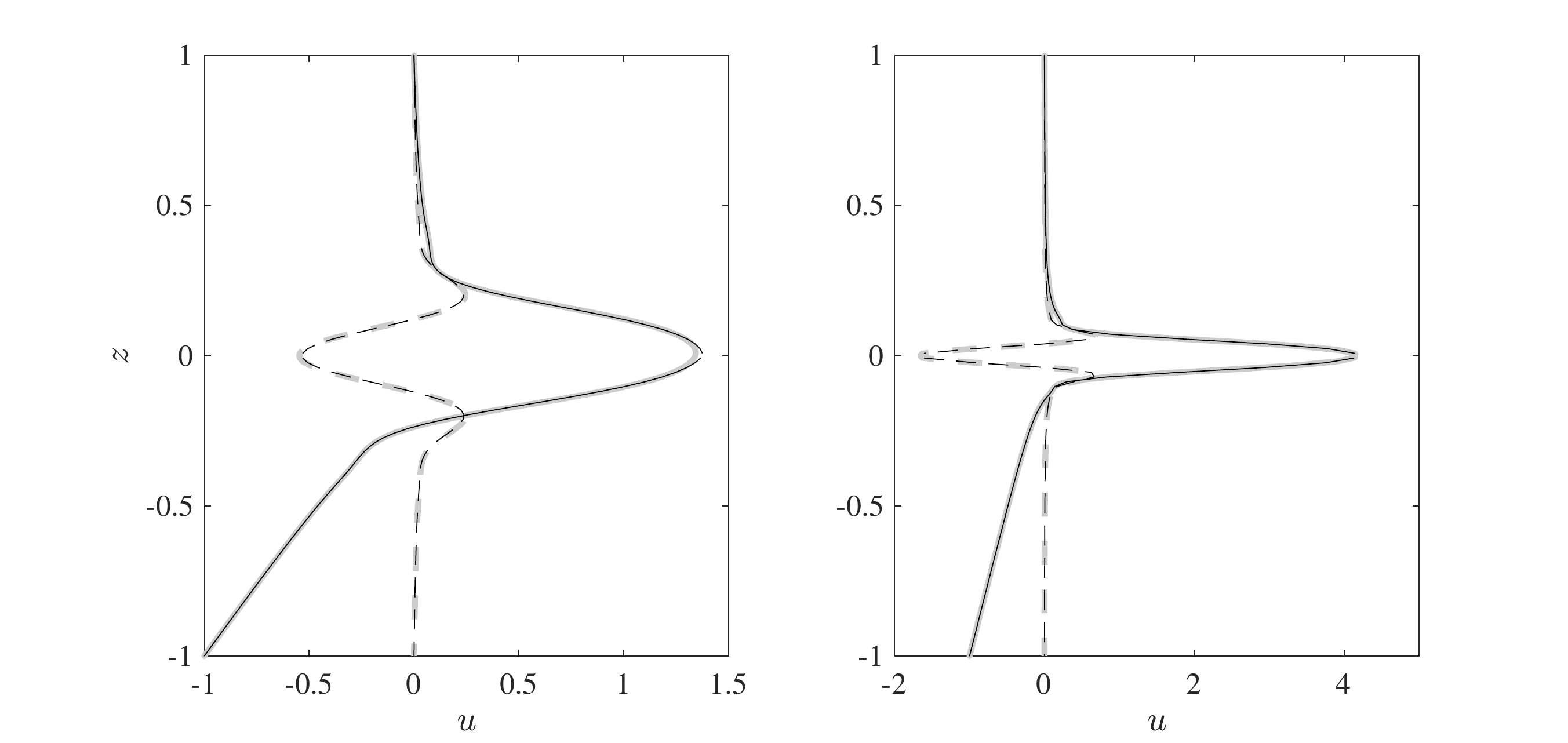}
    \includegraphics[width=\textwidth]{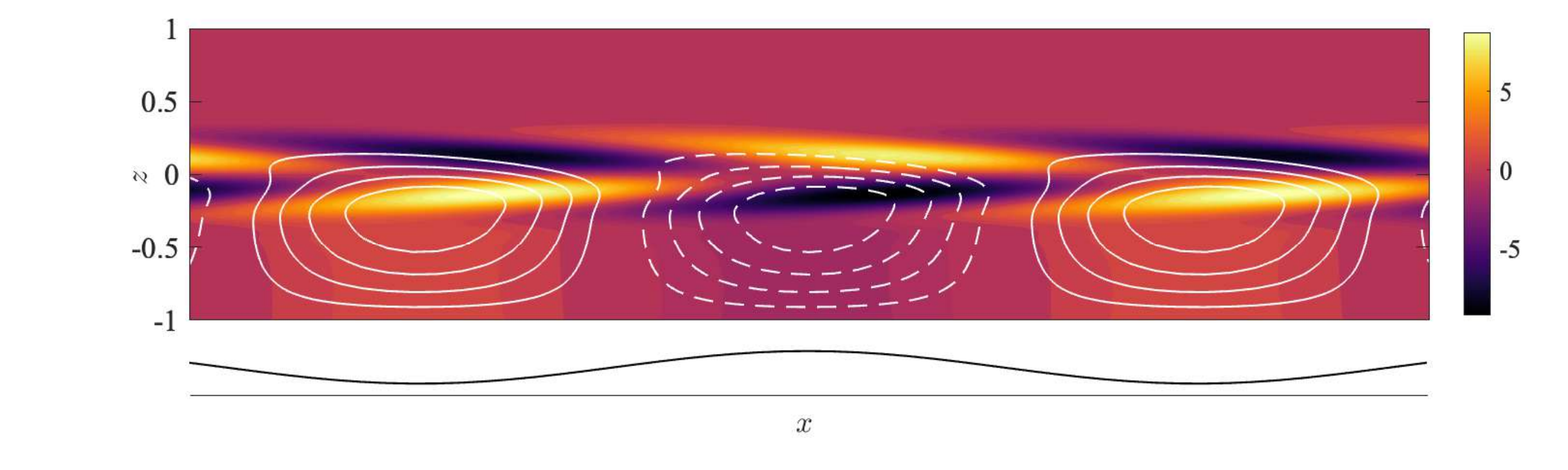}
    \caption{Composite (black) and numerical (grey) solution in the shallow elastic regime. (Top left) $\varepsilon=10^{-4}$. (Top right) $\varepsilon=10^{-6}$. In both cases $R=1$ and $\beta=0.8$. Solid and dashed lines indicate the real and imaginary components of the solution respectively.
    (Bottom) The vorticity field (colours) and streamfunction (lines) corresponding to the $\varepsilon=10^{-4}$ case.}
    \label{fig:se_comp}
\end{figure}

A comparison between the composite solution generated from the above matched asymptotic expansion for the streamwise velocity, $u$, and the numerical solution of the shallow elastic system is provided in figure \ref{fig:se_comp}. 
Excellent agreement is observed.

The solution allows us to estimate the vorticity amplitude in the critical layer, since $\omega = -\dd_z u$ in the shallow approximation. 
Since $u\sim 1/\delta$ and the critical layer is of thickness $\delta$, we have $\omega = O(\delta^{-2}) = O(\sqrt{\alpha W})$.
This scaling is the same as found at the upper wall in a shallow-elastic Couette flow \citep{Page2016}. 

In summary, strongly elastic fluids in shallow channels with long-wave wall undulations experience a significant vorticity amplification at the channel centreline (see bottom panel in figure \ref{fig:se_comp}). 
The vanishing base state stress provides a blocking effect in which the vortices above the surface undulations are restricted to the lower half of the domain.
The rapid drop in vertical velocity leads to a strong jet-like response in the streamwise velocity in a layer of thickness $(\alpha W)^{-1/4}$, which is the source of the vorticity fluctuations. 

\subsection{The shallow elasto-inertial regime}
\label{sec:shallow_ei}
The shallow elasto-inertial regime is associated with significant vorticity amplification at a pair of critical layers in the bulk of the fluid (see figure \ref{fig:mono_regimes}). 
This behaviour requires the influence of both inertia and elasticity, and hence we assume that $E=W/R=O(1)$ here. 
Therefore, the long-wave equations must be modified to account for the fact that pressure scales with the inertial terms. 
In the shallow elasto-inertial regime, the long-wave scaling is,
\begin{subequations}
    \begin{align}
        \hat{u} = u, \quad
        \hat{w} &= \alpha w, \quad
        \hat{p} = p, \\
        \hat{\tau}_{11} = W\tau_{11}, \quad
        \hat{\tau}_{13} &= \alpha W\tau_{13}, \quad
        \hat{\tau}_{33} = \alpha \tau_{33}.
    \end{align}
\end{subequations}
The leading order long-wave equations are now,
\begin{subequations}
    \begin{align}
        \zi u + Dw = 0, \\
        \zi U u + U'w = -\zi p + 
        \varepsilon \beta E D^2 u + (1-\beta)E\left(\zi \tau_{11} +D\tau_{13}\right), \\
        0 = -Dp, \\
        \zi U\tau_{11}+w A_{11}' + \varepsilon\tau_{11} = 
        2\zi A_{11}u + 2\varepsilon T_{13}Du + 2U'\tau_{13}, \\
        \zi U\tau_{13}+\varepsilon w T_{13}' + \varepsilon \tau_{13} = 
        \zi A_{11}w + \varepsilon U'\tau_{33} +\varepsilon^2 Du, \\
        \zi U\tau_{33} + \varepsilon \tau_{33} = 
        2\zi T_{13}w + 2\varepsilon Dw,
    \end{align}
    \label{eqn:sei_full}
\end{subequations}
where as before $A_{11} := T_{11}/W = O(1)$, and the parameter $\varepsilon:=1/(\alpha W)$, which we will subsequently assume to be small.

\begin{figure}
    \centering
    \includegraphics[width=\textwidth]{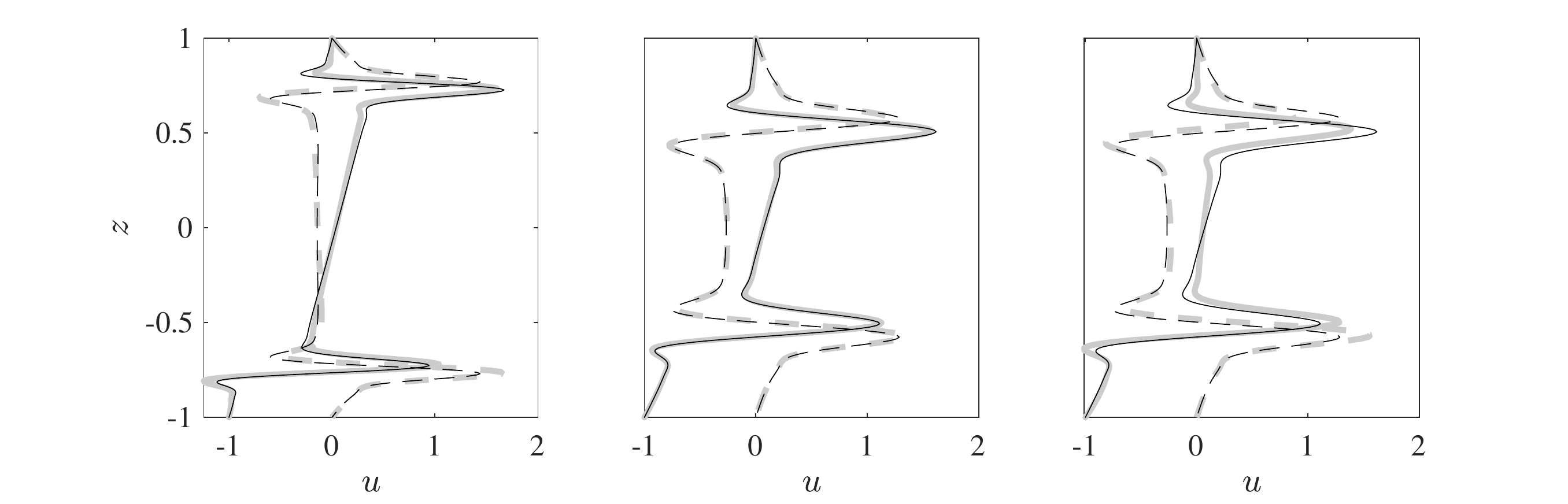}
    \includegraphics[width=\textwidth]{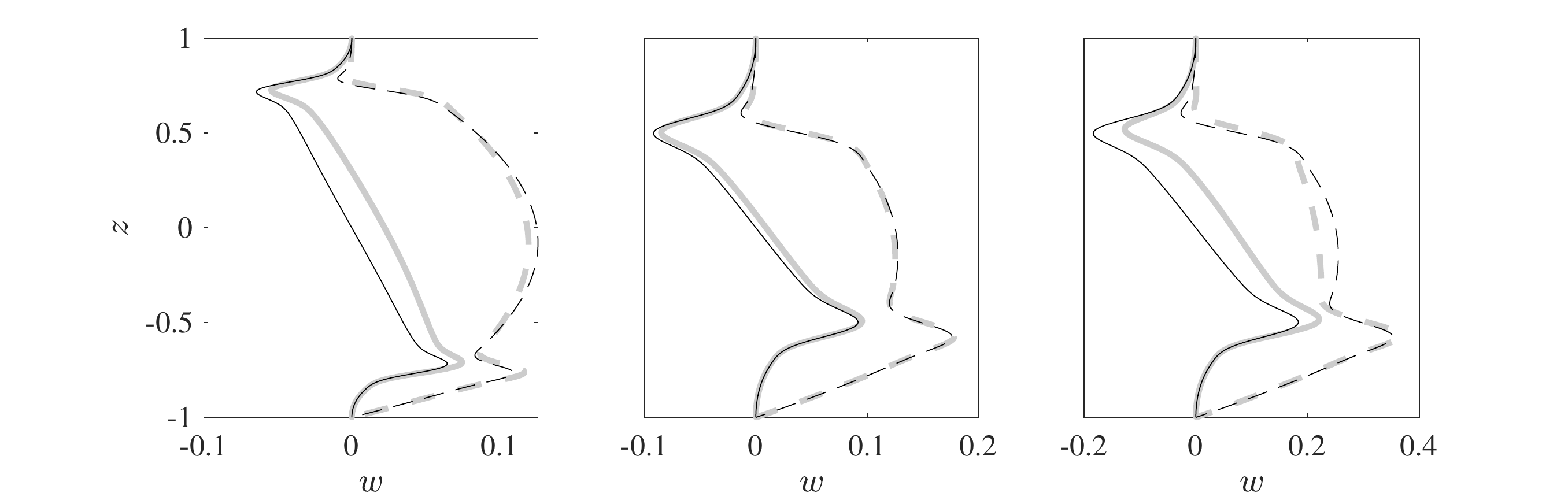}
    \caption{Comparison of full equations (grey) and shallow elasto-inertial approximation (black), with $\varepsilon=10^{-3}$ and $\beta=0.5$. (Left) $\alpha=0.5$, $E=0.1$; (centre) $\alpha=0.5$, $E=0.5$; (right) $\alpha=1$, $E=0.5$. Solid and dashed lines indicate the real and imaginary components of the solution respectively.}
    \label{fig:sei_examples}
\end{figure}
The numerical solution of the elasto-inertial long-wave equations (\ref{eqn:sei_full}) is compared to solution of the original linear system (\ref{eqn:linear_full}) in figure \ref{fig:sei_examples} for modest values of $\alpha$.
Note there are sharp variations in $w$ and enhanced streamwise velocity fluctuations in two thin layers on either side of the channel, which corresponds to the strong vorticity fluctuations observed earlier in figures \ref{fig:local_bump} and \ref{fig:mono_regimes}. 
There is good correspondence between the solution of the long-wave equations and the full numerical solution, even for the relatively large $\alpha=1$.
Therefore, we will again attempt to construct solutions by finding matched asymptotic expansions in $\varepsilon$ within the long-wave system (\ref{eqn:sei_full}) without constructing higher order corrections in $\alpha$.

\emph{Outer solution:}
To leading order in $\varepsilon$ we find the same expressions for the stress as in the shallow elastic regime (see equation \ref{eqn:se_bulk_stress}).
Using these expressions in the streamwise momentum equation and again making use of the streamline displacement, $\phi:=w/U$, we find the long-wave form of the \emph{elastic-Rayleigh} equation \citep{Azaiez1994,Rallison1995,Ray2014},
\begin{equation}
    \left[U^2 - (1-\beta)E A_{11}\right] D\phi_0 = \zi p_0,
\label{eqn:er}
\end{equation}
where the vertical momentum equation indicates again that $p_0=\text{constant}$.
The equation has regular singular points where $(1-\beta)E A_{11}(z) = U^2(z)$.
These are points where the vorticity wave speed, $c_{\omega}(z) = \sqrt{(1-\beta)E A_{11}(z)}$, matches the base flow speed.
The elasto-inertial vorticity wave speed is of the same form as the wavespeed of Alfven waves \citep{chandrasekhar}; here it is the highly tensioned base-flow streamlines that provide a mechanism for wave propagation. 

As noted by \citet{Page2016}, vorticity amplification at the critical layers is associated with a resonance between two fundamental frequencies in the problem:
The singularities in (\ref{eqn:er}) are points in the flow where an observer travelling at the base-flow velocity would sees the wavy wall oscillating at a frequency which is equal to the natural frequency of the elasto-inertial waves. 
The key differences in the channel are (i) that there are two points inside the flow domain where the resonance occurs and (ii) the location of the critical layers does not scale simply with elasticity (the location is $z=\sqrt{2E(1-\beta)}$ in Couette flow).   

In total, there are four values of $z$ where the elastic-Rayleigh equation has singularities, two inside the flow domain where the base-flow velocity cancels with the velocity of backward propagating elasto-inertial waves, $z=\pm\sqrt{\xi_-}$, 
and two outside the flow domain, $z=\pm\sqrt{\xi_+}$, where the base-flow velocity cancels with the forward propagating wavespeed. 
In these expressions,
\begin{equation}
    \xi_{\pm} = \frac{B}{2} \pm \frac{1}{2}\sqrt{B^2 - 4}, \quad B:=2+8E(1-\beta). \nonumber
\end{equation}
Note that the critical layer's distance from the wavy wall can be written in the following form (we have used the lower critical layer here),
\begin{equation}
    l(E) = \left(2(1-\beta)E\right)^{1/2} \, f(\chi),
\end{equation}
where $\chi:=(2 (1-\beta) E)^{-1}$, and $f(\chi) = 1 + \chi - \sqrt{1+\chi^2}$. 
In the low elasticity limit, $E\ll 1$, the critical layer depth scales linearly with the vorticity wave speed based on the wall shear rate, $l\sim \sqrt{2(1-\beta)E}(1 - \sqrt{2(1-\beta)E}\,/2 +\cdots )$, which is a recovery of the result in Couette flow \citep{Page2016}.
In the opposite limit, $E \gg 1$, the critical layer approaches the channel centreline, $l \sim 1 - (2\sqrt{2(1-\beta)E})^{-1} + \cdots$. 
This limit is not physically relevant to the elasto-inertial regime of interest here, since the scalings used to derive (\ref{eqn:sei_full}) no longer hold. Instead, the approach of the critical layer to the centreline signals a transition to the shallow elastic regime which was examined in \S\ref{sec:shallow_elastic}.

The solution for the leading order streamline displacement can be written 
\begin{equation}
    \phi_0^k(z) = C_0^k + \frac{2 \zi p_0 \xi_-}{\xi_-^2 - 1}\left(\Phi(z;\xi_+) - \frac{1}{\sqrt{\xi_-}}\text{log}\left|\frac{\sqrt{\xi_-} - z}{\sqrt{\xi_-} + z}\right|\right).
    \label{eqn:sei_bulk_phi}
\end{equation}
where $\Phi(z;\xi_+) := (1/\sqrt{\xi_+})\text{log}|(\sqrt{\xi_+} - z)/(\sqrt{\xi_+} + z)|$ is analytic in $z\in[-1,1]$, and the superscript $k$ is used to identify the `wall' ($-1\leq z < \sqrt{\xi_-}$), `bulk' ($-\sqrt{\xi_-}<z<\sqrt{\xi_-}$ and `top' ($\sqrt{\xi_-}<z \leq 1$) layers respectively.
We are free to avoid specifying a branch of the logarithm, since phase jumps can be absorbed into the unknown constants $C^k_0$.

For a complete solution in this regime, we will need to match the singular outer solution (\ref{eqn:sei_bulk_phi}) across both critical layers.
We will describe only the procedure at the lower critical layer in detail, noting aspects of the solution that change at the upper layer. 
At the lower layer, we introduce the inner coordinate, $\eta := (z + \sqrt{\xi_-})/\delta$, where the boundary layer thickness $\delta(\varepsilon)$ has not yet been specified. 
Assuming $\eta =O(1)$, the outer vertical velocity has the form,
\begin{align}
    w_0(\eta) \sim &\frac{(1 - \xi_-)C_0^k}{2} + \nonumber \\ 
    &\frac{\zi p_0}{1+\xi_-}\left(\Phi(-\sqrt{\xi_-}) - \frac{1}{\sqrt{\xi_-}}\text{log}|2\sqrt{\xi_-}| + \frac{1}{\sqrt{\xi_-}}\left(\text{log}\delta + \text{log}|\eta|\right)\right) + O(\delta).
    \label{eqn:inner_lim_w}
\end{align}
This suggests that the inner solution will consist of a constant $O(\text{log}\delta)$ vertical velocity and an $O(1)$ contribution that jumps $\jump{w_0} = (1/2)(1-\xi_-)(C_0^b - C_0^w)$ over the critical layer, a jump that must be balanced by large $O(1/\delta)$ streamwise velocity fluctuations.
Unlike the shallow elastic regime, the jump is also associated with significant stress fluctuations, with $\tau_{11}\sim 1/\delta$ as $z\to -\sqrt{\xi_-}$. 

\emph{Inner expansion:}
The appropriate inner scaling in the shallow elasto-inertial regime is 
\begin{subequations}
    \begin{align}
        \overline{u} = \delta u, \quad
        \overline{w} &= w, \quad
        \overline{p} = p, \\
        \overline{\tau}_{11} = \delta\tau_{11}, \quad 
        \overline{\tau}_{13} &= \tau_{13}, \quad
        \overline{\tau}_{33} = \tau_{33}.
    \end{align}
    \label{eqn:inner_var_scales}
\end{subequations}
The $O(\log\delta)$ constant vertical velocity will emerge naturally in the outer limit $\eta \to \pm \infty$. 
We rescale variables following (\ref{eqn:inner_var_scales}) and rewrite our governing equations in terms of the scaled wall-normal variable $\eta:=(z+\sqrt{\xi_-})/\delta$.
Base-flow variables are expanded $U(\eta) = \overline{U}(-\sqrt{\xi_-}) + \overline{U}'(-\sqrt{\xi_-})\delta \eta + \cdots$ etc. 
In these expansions, the bar over the base flow quantities is to emphasise that they are evaluated at the singular point, $z=-\sqrt{\xi_-}$, and are constant.

Assuming $\delta \gg \varepsilon$, approximations to the streamwise and shear polymer stresses in the vicinity of the critical layer are
\begin{subequations}
\begin{align}
    \overline{\tau}_{11} &\sim \frac{1}{\zi \overline{U}}\left(1 - \frac{\overline{U}}{\overline{U}'}\delta \eta\right)\left(2\zi \overline{A}_{11}\overline{u} + 2\zi \overline{A}_{11}'\delta \eta \overline{u} - \delta \overline{A}_{11}'\overline{w}  + \frac{2\overline{U}'\overline{A}_{11}}{\overline{U}}\delta \overline{w}\right) + \cdots, \\
    \overline{\tau}_{13} &\sim \frac{\overline{A}_{11}}{\overline{U}}\overline{w}+ \frac{1}{\overline{U}}\left(\overline{A}_{11}' - \frac{\overline{U}'\overline{A}_{11}}{\overline{U}}\right)\delta \eta \overline{w} + \cdots.
\end{align}
\end{subequations}
Using these approximations in the streamwise momentum equation, along with the fact that $\overline{U}^2 = (1-\beta)E\overline{A}_{11}$, results in 
\begin{equation}
    \underbrace{\left(2\overline{U}' - \frac{(1-\beta)E\overline{A}_{11}'}{\overline{U}}\right)}_{=:\zeta}\zi \eta \overline{u} = -\zi \overline{p} + \frac{\beta E \varepsilon}{\delta^3} \frac{\dd^2 \overline{u}}{\dd \eta^2} + O(\delta).
\end{equation}
A dominant balance indicates that $\delta = (\varepsilon \beta E)^{1/3}(=(\beta/(\alpha R))^{1/3})$ -- the thickness is set by the solvent diffusion lengthscale. 

At leading order, the inner equation for the streamwise velocity at the lower critical layer is 
\begin{equation}
    \frac{\dd^2 \overline{u}_0}{\dd \eta^2} - \zi \zeta \eta \overline{u}_0 = \zi p_0,
\label{eqn:sei_lower_cl}
\end{equation}
where we have used the fact that $\overline{p}_0=p_0$ is constant over the channel depth.
A similar approach at the upper critical layer, $z=+\sqrt{\xi_-}$ leads to an inner equation of the form
\begin{equation}
     \frac{\dd^2 \overline{\overline{u}}_0}{\dd \eta^2} + \zi \zeta \eta \overline{\overline{u}}_0 = \zi p_0,
\end{equation}
the only difference from the lower critical layer being a change in sign on the second term. 
The solution in the upper critical layer can therefore be obtained from a reflection of the lower layer, $\overline{\overline{u}}_0(\eta) = \overline{u}_0(-\eta)$.
A solution to (\ref{eqn:sei_lower_cl}) in terms of Airy functions is provided in Appendix \ref{sec:app_ei}.
For matching, the far-field expression for the inner streamwise velocity is found to be
\begin{equation}
    \overline{u}_0(\eta\to\pm \infty) \sim -\frac{p_0}{\zeta \eta} + \cdots.
\end{equation}
Similar to the shallow elastic regime, the inner solution is directly proportional to the pressure, and can be written in the form $\overline{u}_0= p_0 \Theta(\eta)$.

To match to the outer solution, we integrate with respect to $\eta$ to obtain an expression for inner vertical velocity. 
However, unlike the shallow elastic regime, we are not matching to a constant, but to a logarithmic term. 
Therefore, to avoid divergent integrals, we integrate from some lower limit $-\sigma$ (to be specified) and consider the outer limit of our inner equation with $\sigma \gg |\eta| \gg 1$ above and below the critical layer \citep{asymptotics}.
The vertical velocity is expressed as
\begin{equation}
    \overline{w}_0(\eta) = \tilde{w}_{\sigma} -\zi \int_{-\sigma}^{\eta}\overline{u}_0(\eta')\dd \eta',
\end{equation}
where the constant $\tilde{w}_{\sigma}$ depends on the choice of lower limit $\sigma$.

We now adopt the same approach as in the shallow elastic regime, using the approximate form of $\overline{u}_0$ in the far-field to find asymptotic approximations to $\overline{w}_0$,
\begin{subequations}
\begin{align}
    \overline{w}_0(\eta \ll 1) &\sim \tilde{w}_{\sigma} - \frac{\zi \overline{p}_0}{\zeta}\text{log}|\sigma| + \frac{\zi\overline{p}_0}{\zeta}\text{log}|\eta| + \cdots \\
    \overline{w}_0(\eta \gg 1) &\sim \tilde{w}_{\sigma} - \zi \int_{-\sigma}^{\sigma}\overline{u}_0(\eta')\dd \eta' - \frac{\zi \overline{p}_0}{\zeta}\text{log}|\sigma| + \frac{\zi\overline{p}_0}{\zeta}\text{log}|\eta| + \cdots.
\end{align}
\end{subequations}
Comparing with the inner limit of the outer expansion (\ref{eqn:inner_lim_w}), we see that the choice $\sigma = 1/\delta$ leads to the required $\text{log}\delta$ term for the matching. The integral $\int_{-\sigma}^{\sigma}\overline{u}_0\dd \eta'$ is independent of $\sigma$ when $\sigma \gg 1$ and in fact taking the limit $\sigma\to\infty$ in this integral is convergent due to symmetry.

The outer limits of the inner vertical velocity indicates a jump $\jump{\overline{w}_0} = -\zi\overline{p}_0\int_{-\infty}^{\infty}\Theta \dd \eta'$, which can be compared with predicted outer jumps of $w_0$ to connect the various constants $C_0^k$ across the two critical layers,
\begin{align*}
    C_0^b &= C_0^w - \frac{2\zi p_0}{1-\xi_-}\int_{-\infty}^{\infty}\Theta(\eta')\dd \eta', \\
    C_0^t &= C_0^b - \frac{2\zi p_0}{1-\xi_-}\int_{-\infty}^{\infty}\Theta(\eta')\dd \eta',
\end{align*}
where we have also used the symmetry at the upper critical layer, $\overline{\overline{u}}_0(\eta) = \overline{u}_0(-\eta)$.
Finally, we apply the boundary conditions on the streamline displacement, $\phi_0(-1) = \zi h$ and $\phi_0(1)=0$, which allows the pressure $p_0$ to be written in terms of the roughness height, $h$,
\begin{align}
    -\frac{2\zi p_0 \xi_-}{1-\xi_-^2}\left(\Phi(1) - \frac{1}{\sqrt{\xi_-}}\text{log}\left|\frac{\sqrt{\xi_-} - 1}{\sqrt{\xi_-} + 1}\right|\right) = \zi h &-\frac{2\zi p_0 \xi_-}{1-\xi_-^2}\left(\Phi(-1) - \frac{1}{\sqrt{\xi_-}}\text{log}\left|\frac{\sqrt{\xi_-} + 1}{\sqrt{\xi_-} - 1}\right|\right) \nonumber \\
    &- \frac{4\zi p_0}{1-\xi_-}\int_{-\infty}^{\infty}\Theta(\eta')\dd \eta'.
    \label{eqn:sei_p0}
\end{align}
\begin{figure}
    \centering
    \includegraphics[width=\textwidth]{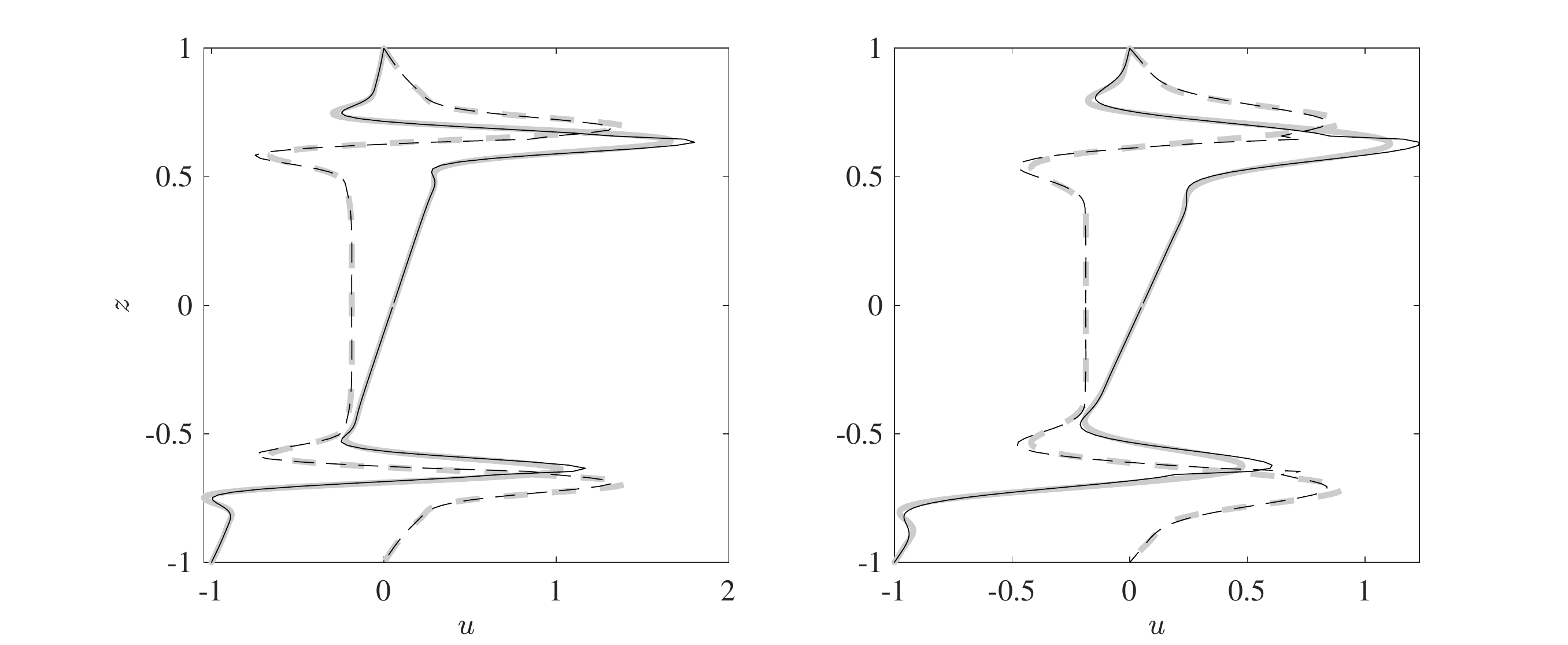}
    \includegraphics[width=\textwidth]{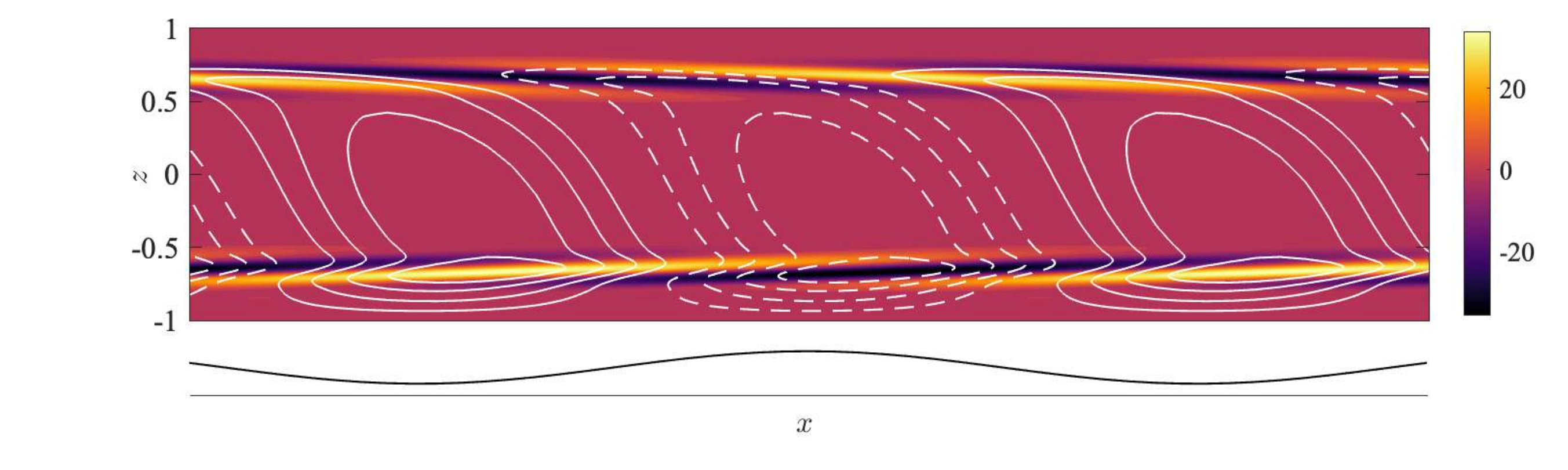}
    \caption{Composite (black) and numerical (grey) solution in the shallow elasto-inertial regime. (Top left) $E=0.2$, $\varepsilon=10^{-3}$, $\beta=0.5$. (Top right) $E=0.5$, $\varepsilon=10^{-3}$, $\beta=0.8$.  Solid and dashed lines indicate the real and imaginary components of the solution respectively.
    (Bottom) The vorticity field (colours) and streamfunction (lines) corresponding to the $E=0.2$ case.}
    \label{fig:sei_comp}
\end{figure}
Rearranging for $p_0$ completes the solution, and a composite solution for $u$ built from the inner/outer asymptotic approximations is compared to the full solution of the shallow system (\ref{eqn:sei_full}) in figure \ref{fig:sei_comp}.
Excellent agreement is observed. 

\begin{figure}
    \centering
    \includegraphics[width=0.4\textwidth]{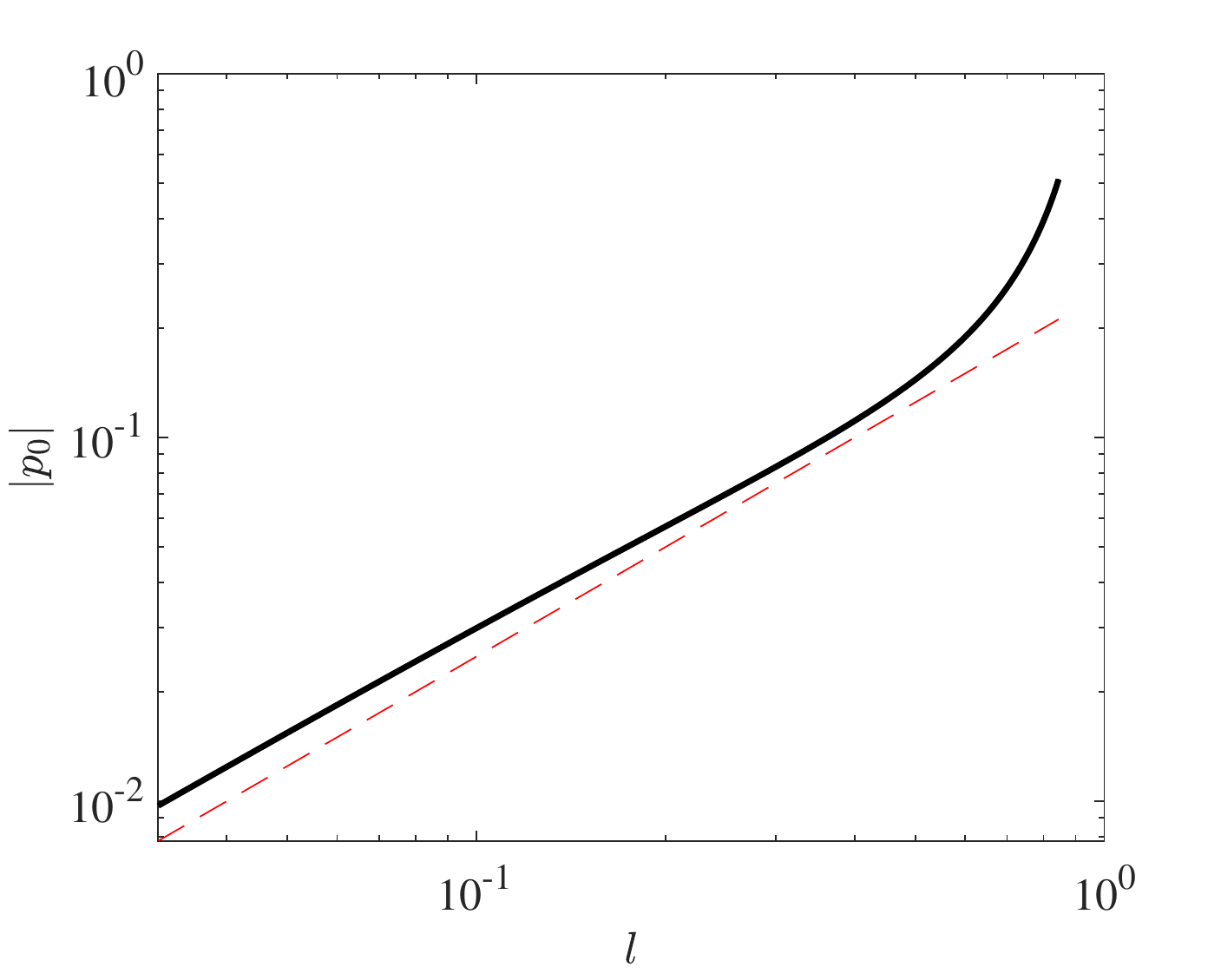}
    \includegraphics[width=0.4\textwidth]{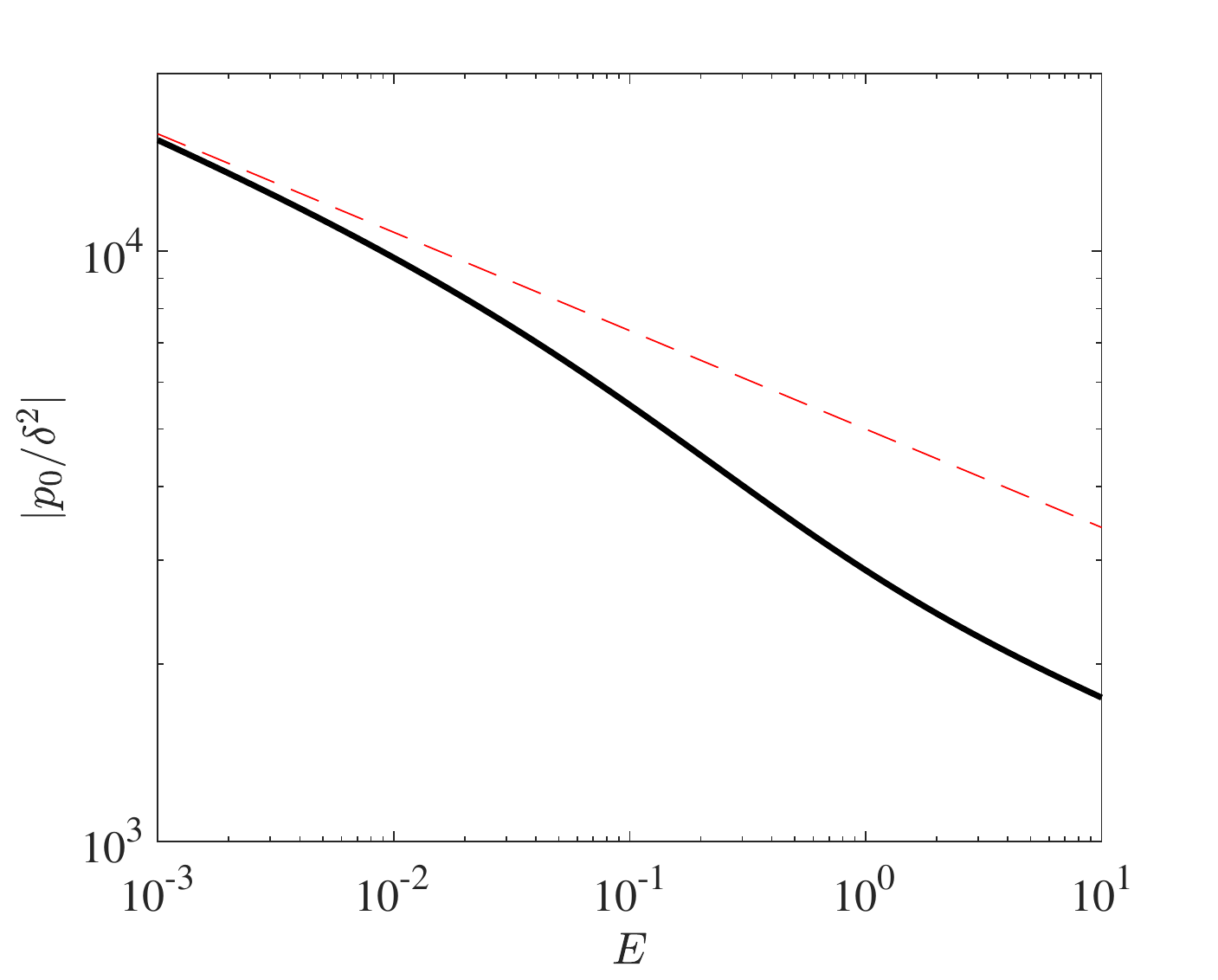}
    \caption{Pressure and vorticity amplification as a function of elasticity, obtained from calculations at $\varepsilon=10^{-6}$, $\beta=0.5$. The red lines indicate (left) $p\sim l(E) = \sqrt{E}$, and (right) $\omega \sim E^{-1/6}$.}
    \label{fig:sei_p}
\end{figure}
Similar to the shallow elastic regime in \S\ref{sec:shallow_elastic}, the vorticity amplification can be estimated from the critical layer scalings, $\omega \sim p_0/\delta^2$. 
However, note that there is a subtle dependence on the elasticity, since $p_0$ is a function of $E$ via the location of the critical layers ($z=\pm\sqrt{\xi_-}$) in equation (\ref{eqn:sei_p0}) and the boundary-layer thickness is itself dependent on $E$ indirectly if we are examining the scaling as a function of $\varepsilon$. 

We examine the dependence of the bulk pressure $p_0$ and the vorticity on the elasticity numerically in figure \ref{fig:sei_p}.
The pressure scales linearly with the critical layer distance from the lower wall $p_0\sim l$, which for small elasticity $l\approx\sqrt{2(1-\beta)E}$ (see earlier discussion around the elastic-Rayleigh equation \ref{eqn:er}).
For fixed $\varepsilon$, the boundary layer thickness $\delta \propto E^{1/3}$, which indicates the $\omega \sim E^{-1/6}/\varepsilon^{2/3}$, a scaling which is confirmed in figure \ref{fig:sei_p} for a wide range of elasticities. 

In summary, the shallow elasto-inertial regime is associated with a significant amplification of spanwise vorticity at two critical layers, where the base velocity matches an elasto-inertial wavespeed (see bottom panel in figure \ref{fig:sei_comp}). 
The thickness of these layers is set by a diffusion lengthscale in the solvent, $\delta \sim (\alpha R/\beta)^{-1/3}$.
The vorticity amplification scales with $1/\delta^2$, but has a more complex dependence on the elasticity which is fixed by the bulk pressure, as detailed above. 
Significantly, the response in the upper critical layer is equal in magnitude to that at the lower layer, hence the perturbation vorticity fills the depth of the channel. 

\subsection{Discussion}
\label{sec:discussion}
\begin{figure}
    \centering
    \includegraphics[width=0.9\textwidth]{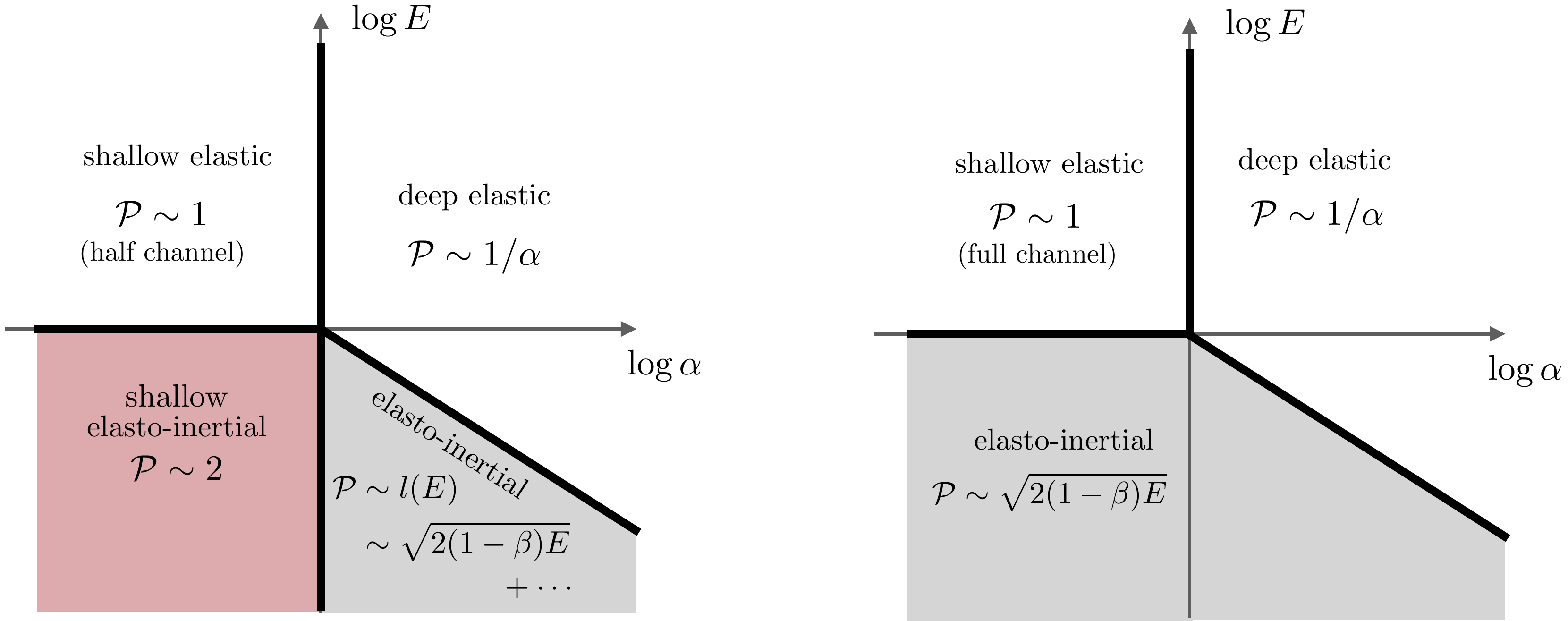}
    \caption{Phase diagrams for the wavy channel (left) and also wavy Couette flow from \citet{Page2016} (right). Also included is an estimate of the penetration depth, $\mathcal P$, of the vorticity perturbation induced at the lower wall. Note that the diagonal line in the bottom right quadrant indicates $1/\alpha \sim \sqrt{2(1-\beta)E}$.}
    \label{fig:phase_diagram}
\end{figure}
In the asymptotic solutions above we have demonstrated how the change in geometry from Couette to channel flow has a substantial impact on the vorticity field generated by surface roughness on the lower channel wall. 
These differences only manifest when the channel depth is smaller than the roughness lengthscale; otherwise the perturbation is exposed to a simple shear (to a leading approximation) before it has decayed -- and therefore we did not study this case here \citep[see the Couette analysis in][]{Page2016}. 
However, when the channel is `shallow' ($\alpha \lesssim 1$) the non-monatonic velocity profile leads to two striking new behaviours: shallow elastic flow (a regime also found in Couette but with very different phenomenology) and shallow elasto-inertial flow (not observed in Couette at all).

Some of these differences are summarized in the phase diagrams reported in figure \ref{fig:phase_diagram}, where the Couette results have also been included for reference. 
In this image, the various regimes are classified by the values of surface wavenumber, $\alpha$, and elasticity, $E$ (assuming that the Weissenberg number is large); the `shallow' geometries studied in detail in this paper are to the left of the vertical $\alpha=1$ line.
Note that the diagrams are rotated relative to the one previously reported in \citet{Page2016} due to the use of the channel half-height as the lengthscale, rather than the inverse wavenumber of the surface wave. 
The diagonal line identifies the transisition from elasto-inertial to deep elastic flow when the location of the (lower) critical layer is further than a surface wavelength from the lower wall, $\sqrt{2(1-\beta)E}>1/\alpha$ \citep[see][for details of these regimes which occur in deep channels, and are unchanged from the Couette results]{Page2016}.

An estimate of the vorticity penetration depth, $\mathcal P$, is also provided for each regime in figure \ref{fig:phase_diagram}. 
This a measure of how far the perturbation vorticity penetrates into the bulk of the flow above the wavy boundary.
There are various ways to define this quantity, for instance in \citet{Page2016} an integral measure is used, and $\mathcal P$ is defined as the $z$-location below which 99\% of the total perturbation enstrophy is contained. 
In shallow elastic flow, the vorticity amplification mechanism is essentially due to conservation of mass, whereby disconnected purely elastic responses in either half of the channel drive a discontinuity in vertical velocity, which leads to a strong streamwise velocity response at the centreline in a layer of thickness $\delta \sim (\alpha W)^{-1/4}$.
The penetration depth of the vorticity is therefore approximately the channel half-height, $\mathcal P\sim 1$, though for the boundary layer thickness shrinks slowly with increasing Weissenberg number, so a more refined estimate would be $\mathcal P \sim 1 + \delta$. 

In shallow elasto-inertial flow, the same resonance between elastic waves and the frequency of wall oscillations that was found in the Couette flow occurs again, but now at a pair of critical layers. 
Consequently, a strong vorticity perturbation is generated over the entire channel depth -- which should be contrasted to the Couette flow where it is limited to the single critical layer (i.e. near the lower wall only). 
As such, the penetration depth in shallow elasto-inertial flows is the full channel depth $\mathcal P \sim 2$, while in a Couette flow there is no distinction between shallow and deep elasto-inertial geometries, since the vorticity penetration depth is set by the height of the single critical layer, $\mathcal P \sim \sqrt{2(1-\beta)E}$.  

Finally, we comment on the transition from shallow elasto-inertial to shallow elastic flows which can be initiated at fixed $\alpha<1$ by increasing the elasticity and is identified by the thick horizontal black line on $E=1$ in figure \ref{fig:phase_diagram}. 
In the Couette flow \citep{Page2016}, the transition between shallow elastic and elasto-inertial behaviour is relatively simple: it occurs when the critical layer height crosses the top boundary. 
The transition is less straightforward in the channel for two reasons:
(1) the critical layers never leave the flow domain, but instead both move towards the centre of the channel as the elasticity is increased (see equation \ref{eqn:sei_lower_cl}) -- and too large an elasticity breaks the dominant balance assumed in the elasto-inertial flow; (2) the thickness of the boundary layer at $z=0$ in shallow elastic flow is relatively large, $\delta \sim (\alpha W)^{-1/4}$, and shrinks very slowly with increasing $W$.
As such, for a wide range of flow parameters it is possible to find a response which shares some characteristics of both regimes. 
For instance, this can be seen for the `intermediate' states reported in figures \ref{fig:local_bump} and \ref{fig:mono_regimes}. 

\section{Conclusion}
\label{sec:conclusions}
In this paper, we have shown how significant vorticity fluctuations can be generated by long-wave surface undulations in planar viscoelastic channel flows.
Unlike Newtonian flows in the same configuration, the response to the surface displacement can be significant across a full range of Reynolds numbers.
The new regimes of vortical penetration -- shallow elastic and shallow elasto-inertial flows -- are both primarily associated with the vertical dependence of the base-state streamwise normal stress, $T_{11}(z)$, and its symmetry about the midplane. 
The vanishing stress at $z=0$ confines inertialess, high-elastic responses to the lower half of the channel, while the symmetry in elasto-inertial flows drives a response at two critical layers in the bulk of the flow. 


Future studies should address the influence of finite amplitude wall roughness on the behaviours discovered in this work. 
For instance, in the shallow elastic flow there is a strong asymmetry in the response, where the perturbation vortices are confined to the lower half of the channel. It would be of interest to explore at what roughness amplitude this has an appreciable impact on the velocity and conformation profiles, which can be explored with the tools developed by \citet{hameduddin_meneveau_zaki_gayme_2018,hameduddin_zaki_2019}. 
The secondary instability of these flows are also of significant interest owing to the possibility for non-trivial dynamics at low-$R$ -- for instance are the instabilities three dimensional \citep[e.g. like elastic `Taylor' rolls, see][]{Larson1990} or two-dimensional leading to structures observed in flows with moderate inertia \citep{Page2020}. 
It would also be of interest to explore how some of the effects considered in this paper change for other non-monotonic background stress fields (e.g. in boundary layers).

\section*{Acknowledgements}
JP acknowledges support from EPSRC under grant number EP/V027247/1, and TAZ acknowledges support from the National Science Foundation (NSF) Grant No. CBET-2027875.

\vspace{1em}
Declaration of Interests. The authors report no conflict of interest.

\appendix
\section{Details of the inner solutions}
\subsection{The shallow elastic regime}
\label{sec:app_se}
In the shallow elastic regime there is a boundary layer at $z=0$, where a dominant balance between solvent diffusion and the polymer force in the streamwise momentum equation yields 
\begin{equation}
    \frac{\dd^2 \overline{u}_0}{\dd \eta^2} + g(\beta) \eta^2 \overline{u}_0 = \frac{\zi R p_1}{\beta},
    \label{eqn:inner_u0_app}
\end{equation}
where $g(\beta):=4\zi(1-\beta)/\beta$.
Note there is a reflection symmetry in the equation and matching conditions, hence $\overline{u}_0(-\eta)=\overline{u}_0(\eta)$.
A numerically satisfactory pair of solutions to equation (\ref{eqn:inner_u0_app}) is 
\begin{align}
    \mathscr U_1(\eta) &= (\eta^2)^{1/4}I_{-1/4}(\sqrt{-g}\eta^2/2) + \frac{\eta}{(\eta^2)^{1/4}}I_{1/4}(\sqrt{-g}\eta^2/2), \nonumber \\
    \mathscr U_2(\eta) &= (\eta^2)^{1/4}I_{-1/4}(\sqrt{-g}\eta^2/2) - \frac{\eta}{(\eta^2)^{1/4}}I_{1/4}(\sqrt{-g}\eta^2/2),
\end{align}
where the $I_k$ are modified Bessel functions of the first kind.
The Wronskian is $W=-4\sqrt{2}/\pi$.

Using variation of parameters, we write the complete solution to the inner equation as
\begin{equation}
    \overline{u}_0(\eta) = \overline{C}_1\mathscr U_1(\eta) + \overline{C}_2\mathscr U_2(\eta) -\frac{\zi R\pi p_1}{4\sqrt{2}\beta}\left(\mathscr U_2\int_0^{\eta}\mathscr U_1(\eta')\dd \eta' - \mathscr U_1\int_0^{\eta}\mathscr U_2(\eta')\dd \eta'\right).
\end{equation}
Applying boundedness as $\eta \to \pm \infty$, we find
\begin{align*}
    \overline{C}_1 &= -\frac{\zi \pi R p_1}{4\sqrt{2}\beta}\int_{0}^{\infty}\mathscr U_2(\eta')\dd \eta', \\
    \overline{C}_2 &= -\frac{\zi \pi R p_1}{4\sqrt{2}\beta}\int_{-\infty}^{0}\mathscr U_1(\eta')\dd \eta'.
\end{align*}
The inner solution can therefore be written in the form $\overline{u}_0(\eta) = p_1\Theta(\eta)$, where the function $\Theta$ is independent of the pressure. 

To obtain the asymptotic form of the inner streamwise velocity in the far field (i.e. as $\eta\to\pm\infty$), we need large-$\eta$ asymptotic approximations to integrals of the form $\int_0^{\eta}\mathscr U_j(\eta')\dd \eta'$.
These limiting forms are straightforward to derive by writing $\mathscr U_j = -(g\eta^2)^{-1} \dd^2_{\eta}\mathscr U_j$ in the integrand and using repeated integration by parts \citep[e.g. see][]{asymptotics}.
We find,
\begin{equation}
    \overline{u}_0(\eta \to \pm\infty) \sim \frac{R p_1}{4(1-\beta)}\left(\frac{1}{\eta^2} - \frac{6}{g\eta^6} + \cdots \right).
\end{equation}

\subsection{The shallow elasto-inertial regime}
\label{sec:app_ei}
In the shallow elasto-inertial regime there are critical layers at $z=\pm\sqrt{\xi_-}$ (see discussion in \S\ref{sec:shallow_ei}). 
In the lower critical layer, a balance between solvent diffusion, advection and polymer force reduces the streamwise momentum equation to the form, 
\begin{equation}
    \frac{\dd^2 \overline{u}_0}{\dd \eta^2} - \zi \zeta \eta \overline{u}_0 = \zi p_0,
\end{equation}
where the constant, $\zeta$, can be written in terms of the critical layer location $\zeta \equiv (\xi_- + 1)/\sqrt{\xi_-}$.
This equation is an inhomogeneous Airy equation; a numerically satisfactory pair of solutions is
\begin{align}
    \mathscr U_1(\eta) &= \text{Ai}(e^{\zi \pi/6}\zeta^{1/3}\eta), \nonumber \\
    \mathscr U_2(\eta) &= \text{Ai}(e^{\zi \pi/6}\zeta^{1/3}\eta)- \zi\,\text{Bi}(e^{\zi \pi/6}\zeta^{1/3}\eta),
\end{align}
with Wronskian $W=(-\zi/\pi)e^{\zi\pi/6}\zeta^{1/3}$.
Note that at the upper critical layer, $z=+\sqrt{\xi_-}$, we find the streamwise velocity solves
\begin{equation}
     \frac{\dd^2 \overline{\overline{u}}_0}{\dd \eta^2} + \zi \zeta \eta \overline{\overline{u}}_0 = \zi p_0,
\end{equation}
and so the solution in the upper critical layer can be obtained from a reflection of the lower layer solution, $\overline{\overline{u}}_0(\eta) = \overline{u}_0(-\eta)$. 
Therefore, we discuss only the solution in the lower layer here. 

Using the method of variation of parameters, the inner solution may be written in the form
\begin{equation}
    \overline{u}_0(\eta) = \overline{C}_1\mathscr U_1(\eta) + \overline{C}_2\mathscr U_2(\eta) -\frac{\pi e^{-\zi \pi/6}p_0}{\zeta^{1/3}}\left(\mathscr U_2\int_0^{\eta}\mathscr U_1(\eta')\dd \eta' - \mathscr U_1\int_0^{\eta}\mathscr U_2(\eta')\dd \eta'\right).
\end{equation}
Applying boundedness as $\eta \to \pm \infty$, we find
\begin{align*}
    \overline{C}_1 &= \frac{\pi e^{-\zi \pi/6}p_0}{\zeta^{1/3}}\int_{-\infty}^{0}\mathscr U_2(\eta')\dd \eta', \\
    \overline{C}_2 &= \frac{\pi e^{-\zi \pi/6}p_0}{\zeta^{1/3}}\int_{0}^{\infty}\mathscr U_1(\eta')\dd \eta'.
\end{align*}
Note now that the solution for the inner velocity can be written in the form $\overline{u}_0(\eta) = p_0 \Theta(\eta)$, where $\Theta$ is an $O(1)$ function which is independent of the pressure.

The asymptotic form of $\overline{u}_0(\eta \to \pm \infty)$ is straightforward to derive by using integration by parts in a similar approach to that adopted for the shallow elastic regime.
Only the first term in the expansion is required for the matching with the outer solution, 
\begin{equation}
    \overline{u}_0(\eta\to\pm \infty) \sim -\frac{p_0}{\zeta \eta} + \cdots.
\end{equation}


\bibliographystyle{jfm}

\begin{thebibliography}{31}
\expandafter\ifx\csname natexlab\endcsname\relax\def\natexlab#1{#1}\fi
\bibitem[Azaiez \& Homsy(1994)]{Azaiez1994}
{\sc Azaiez, J. \& Homsy, G.~M.} 1994 {Linear stability of free shear flow of
  viscoelastic liquids}. {\em Journal of Fluid Mechanics\/} {\bf 268}, 37--69.

\bibitem[Bender \& Orszag(1978)]{asymptotics}
{\sc Bender, C.~M. \& Orszag, S.~A.} 1978 {\em {Advanced Mathematical Methods
  for Scientists and Engineers}\/}, 1st edn. Singapore: McGraw-Hill.

\bibitem[Buza {\em et~al.\/}(2021)Buza, Page \& Kerswell]{Buza2021}
{\sc Buza, Gergely, Page, Jacob \& Kerswell, Rich~R.} 2021 Weakly nonlinear
  analysis of the viscoelastic instability in channel flow for finite and
  vanishing reynolds numbers.

\bibitem[Chandrasekhar(1961)]{chandrasekhar}
{\sc Chandrasekhar, S.} 1961 {\em {Hydrodynamic and hydromagnetic
  stability}\/}. New York: Dover.

\bibitem[Charru \& Hinch(2000)]{charru2000}
{\sc Charru, F. \& Hinch, E.~J.} 2000 {`Phase diagram' of interfacial
  instabilities in a two-layer Couette flow and mechanism of the long-wave
  instability}. {\em Journal of Fluid Mechanics\/} {\bf 414}, 195--223.

\bibitem[Dubief {\em et~al.\/}(2013)Dubief, Terrapon \& Soria]{Dubief2013}
{\sc Dubief, Y., Terrapon, V.~E. \& Soria, J.} 2013 {On the mechanism of
  elasto-inertial turbulence.} {\em Physics of Fluids\/} {\bf 25}~(11), 110817.

\bibitem[Garg {\em et~al.\/}(2018)Garg, Chaudhary, Khalid, Shankar \&
  Subramanian]{Garg2018}
{\sc Garg, Piyush, Chaudhary, Indresh, Khalid, Mohammad, Shankar, V. \&
  Subramanian, Ganesh} 2018 Viscoelastic pipe flow is linearly unstable. {\em
  Phys. Rev. Lett.\/} {\bf 121}, 024502.

\bibitem[Groisman \& Steinberg(2000)]{Groisman2000}
{\sc Groisman, A. \& Steinberg, V.} 2000 {Elastic turbulence in a polymer
  solution flow}. {\em Nature\/} {\bf 405}, 53--55.

\bibitem[Hameduddin {\em et~al.\/}(2018)Hameduddin, Meneveau, Zaki \&
  Gayme]{hameduddin_meneveau_zaki_gayme_2018}
{\sc Hameduddin, Ismail, Meneveau, Charles, Zaki, Tamer~A. \& Gayme,
  Dennice~F.} 2018 Geometric decomposition of the conformation tensor in
  viscoelastic turbulence. {\em Journal of Fluid Mechanics\/} {\bf 842},
  395–427.

\bibitem[Hameduddin \& Zaki(2019)]{hameduddin_zaki_2019}
{\sc Hameduddin, Ismail \& Zaki, Tamer~A.} 2019 The mean conformation tensor in
  viscoelastic turbulence. {\em Journal of Fluid Mechanics\/} {\bf 865},
  363–380.

\bibitem[Haward {\em et~al.\/}(2018{\natexlab{{\em a\/}}})Haward, Page, Zaki \&
  Shen]{Haward2018a}
{\sc Haward, Simon~J., Page, Jacob, Zaki, Tamer~A. \& Shen, Amy~Q.}
  2018{\natexlab{{\em a\/}}} Inertioelastic poiseuille flow over a wavy
  surface. {\em Phys. Rev. Fluids\/} {\bf 3}, 091302.

\bibitem[Haward {\em et~al.\/}(2018{\natexlab{{\em b\/}}})Haward, Page, Zaki \&
  Shen]{Haward2018b}
{\sc Haward, Simon~J., Page, Jacob, Zaki, Tamer~A. \& Shen, Amy~Q.}
  2018{\natexlab{{\em b\/}}} “phase diagram” for viscoelastic poiseuille
  flow over a wavy surface. {\em Physics of Fluids\/} {\bf 30}~(11), 113101.

\bibitem[Haward {\em et~al.\/}(2017)Haward, Shen, Page \& Zaki]{Haward2017}
{\sc Haward, Simon~J., Shen, Amy~Q., Page, Jacob \& Zaki, Tamer~A.} 2017
  Poiseuille flow over a wavy surface. {\em Phys. Rev. Fluids\/} {\bf 2},
  124102.

\bibitem[Jovanovic \& Kumar(2010)]{Jovanovic2010}
{\sc Jovanovic, M.~R. \& Kumar, S.} 2010 {Transient growth without inertia}.
  {\em Physics of Fluids\/} {\bf 22}, 023101.

\bibitem[Jovanovic \& Kumar(2011)]{Jovanovic2011}
{\sc Jovanovic, M.~R. \& Kumar, S.} 2011 {Nonmodal amplification of stochastic
  disturbances in strongly elastic channel flows}. {\em Journal of
  Non-Newtonian Fluid Mechanics\/} {\bf 166}, 755--778.

\bibitem[Khalid {\em et~al.\/}(2021)Khalid, Shankar \& Subramanian]{Khalid2021}
{\sc Khalid, Mohammad, Shankar, V. \& Subramanian, Ganesh} 2021 Continuous
  pathway between the elasto-inertial and elastic turbulent states in
  viscoelastic channel flow. {\em Phys. Rev. Lett.\/} {\bf 127}, 134502.

\bibitem[Larson {\em et~al.\/}(1990)Larson, Shaqfeh \& Muller]{Larson1990}
{\sc Larson, R.~G., Shaqfeh, E. S.~G. \& Muller, S.~J.} 1990 {A purely elastic
  instability in Taylor-Couette flow}. {\em Journal of Fluid Mechanics\/} {\bf
  218}, 573--600.

\bibitem[Lee \& Zaki(2017)]{Lee2017}
{\sc Lee, Sang~Jin \& Zaki, Tamer~A.} 2017 Simulations of natural transition in
  viscoelastic channel flow. {\em Journal of Fluid Mechanics\/} {\bf 820},
  232–262.

\bibitem[Meulenbroek {\em et~al.\/}(2004)Meulenbroek, Storm, Morozov \& van
  Saarloos]{Meulenbroek2004}
{\sc Meulenbroek, B., Storm, C., Morozov, A.~N. \& van Saarloos, W.} 2004
  {Weakly nonlinear subcritical instability of visco-elastic Poiseuille flow}.
  {\em Journal of Non-Newtonian Fluid Mechanics\/} {\bf 116}~(2-3), 235--268.

\bibitem[Morozov \& Saarloos(2005)]{Morozov2005}
{\sc Morozov, A. \& Saarloos, W.~V.} 2005 {Subcritical Finite-Amplitude
  Solutions for Plane Couette Flow of Viscoelastic Fluids}. {\em Physical
  Review Letters\/} {\bf 95}~(2), 1--4.

\bibitem[Page {\em et~al.\/}(2020)Page, Dubief \& Kerswell]{Page2020}
{\sc Page, Jacob, Dubief, Yves \& Kerswell, Rich~R.} 2020 Exact traveling wave
  solutions in viscoelastic channel flow. {\em Phys. Rev. Lett.\/} {\bf 125},
  154501.

\bibitem[Page \& Zaki(2015)]{Page2015}
{\sc Page, J. \& Zaki, T.~A.} 2015 {The dynamics of spanwise vorticity
  perturbations in homogeneous viscoelastic shear flow}. {\em Journal of Fluid
  Mechanics\/} {\bf 777}, 327--363.

\bibitem[Page \& Zaki(2016)]{Page2016}
{\sc Page, J. \& Zaki, T.~A.} 2016 {Viscoelastic shear flow a wavy surface}.
  {\em Journal of Fluid Mechanics\/} {\bf 901}, 392--429.

\bibitem[Pan {\em et~al.\/}(2013)Pan, Morozov, Wagner \& Arratia]{Pan2013}
{\sc Pan, L., Morozov, A., Wagner, C. \& Arratia, P.~E.} 2013 {Nonlinear
  elastic instability in channel flows at low Reynolds numbers}. {\em Physical
  Review Letters\/} {\bf 110}, 174502.

\bibitem[Rallison \& Hinch(1995)]{Rallison1995}
{\sc Rallison, J.~M. \& Hinch, E.~J.} 1995 {Instability of a high-speed
  submerged elastic jet}. {\em Journal of Fluid Mechanics\/} {\bf 288},
  311--324.

\bibitem[Ray \& Zaki(2014)]{Ray2014}
{\sc Ray, P.~K. \& Zaki, T.~A.} 2014 {Absolute instability in viscoelastic
  mixing layers}. {\em Physics of Fluids\/} {\bf 26}~(1), 014103.

\bibitem[Samanta {\em et~al.\/}(2013)Samanta, Dubief, Holzner, Sch\"{a}fer,
  Morozov, Wagner \& Hof]{Samanta2013}
{\sc Samanta, D.~S., Dubief, Y., Holzner, H., Sch\"{a}fer, C., Morozov, A.~N.,
  Wagner, C. \& Hof, B.} 2013 {Elasto-inertial turbulence}. {\em Proceedings of
  the National Academy of Sciences of the United States of America\/} {\bf
  110}, 10557--10562.

\bibitem[Shaqfeh(1996)]{Shaqfeh1996}
{\sc Shaqfeh, E. S.~G.} 1996 {Purely Elastic Instabilities in Viscometric
  Flows}. {\em Annual Review of Fluid Mechanics\/} {\bf 28}, 129--185.

\bibitem[Shekar {\em et~al.\/}(2020)Shekar, McMullen, McKeon \&
  Graham]{Shekar2020}
{\sc Shekar, Ashwin, McMullen, Ryan~M., McKeon, Beverley~J. \& Graham,
  Michael~D.} 2020 Self-sustained elastoinertial tollmien–schlichting waves.
  {\em Journal of Fluid Mechanics\/} {\bf 897}, A3.

\bibitem[White \& Mungal(2008)]{White2008}
{\sc White, C.~M. \& Mungal, M.~G.} 2008 {Mechanics and prediction of turbulent
  drag reduction with polymer additives}. {\em Annual Review of Fluid
  Mechanics\/} {\bf 40}, 235--256.

\bibitem[Yoo \& Joseph(1985)]{Yoo1985}
{\sc Yoo, J.~Y. \& Joseph, D.~D.} 1985 {Hyperbolicity and change of type in the
  flow of viscoelastic fluids through channels}. {\em Journal of Non-Newtonian
  Fluid Mechanics\/} {\bf 19}, 15--41.

\end{thebibliography}

\end{document}